\documentclass[12pt,a4paper]{iopart}
\usepackage{amssymb}
\usepackage{amsmath}
\usepackage{graphicx}
\usepackage{color}

\newcommand{\data}{{\scriptscriptstyle {\rm DATA}}}

\newcommand{\brb}[1]{\left[#1\right]}
\newcommand{\brc}[1]{\left<#1\right>}
\newcommand{\bre}[1]{\left\{#1\right\}}

\newcommand{\be}{\begin{equation}}
\newcommand{\ee}{\end{equation}}
\newcommand{\benn}{\begin{equation*}}
\newcommand{\eenn}{\end{equation*}}
\newcommand{\bea}{\begin{eqnarray}}
\newcommand{\eea}{\end{eqnarray}}

\newcommand{\us}{\underline{s}}
\newcommand{\uv}{\underline{v}}

\begin{document}

\title[Data quality for the inverse Ising problem]{Data quality for the inverse Ising problem}

\author{Aur\'elien Decelle$^1$, Federico Ricci-Tersenghi$^2$, Pan Zhang$^3$}

\address{$^1$ Laboratoire de Recherche en Informatique, TAO - INRIA, CNRS, Univ. Paris-Sud, Universit\'e Paris-Saclay, B\^at. 660, 91190 Gif-sur-Yvette, France.}
\address{$^2$ Dipartimento di Fisica, INFN--Sezione di Roma1 and CNR--Nanotec, Universit\`a La Sapienza, Piazzale Aldo Moro 5, I-00185 Roma, Italy.}
\address{$^3$ Institute of Theoretical Physics, Chinese Academy of Sciences, Zhong-Guan-Cun-Dong-Lu 55, Beijing 100190, China.}

\begin{abstract}
	There are many methods proposed for inferring parameters of the Ising 
	model from given data, that is a set of configurations generated according
	to the model itself. However little attention has been paid until now to 
	the data, e.g.\
	how the data is generated, whether the inference error using one set of data could
	be smaller than using another set of data, etc. In this paper we discuss the 
	data quality problem in the inverse Ising problem, using as a benchmark the kinetic Ising model.
	We quantify the quality
	of data using effective rank of the correlation matrix, and show that 
	data gathered in a out-of-equilibrium regime has a better quality than data 
	gathered in equilibrium for coupling reconstruction. We also propose a matrix-perturbation based method
	for tuning the quality of given data and for removing bad-quality (i.e.\ redundant) configurations from data.
\end{abstract}
\submitto{\jpa}

\maketitle

\section{Introduction}

In the past few years, considerable attention has been drawn to the inverse Ising problems that study how to infer or reconstruct the parameters of an Ising model from configurations generated according to the model itself.
This inference process, also known as ``Boltzmann machine learning'' in computer science \cite{ackley1985learning}, is linked to the maximum entropy principle applied to models of pairwise interacting variables when the first two moments of data are measured \cite{Schneidman_etal_2006_nature}, and has a capability to model rich behaviors of observed data. Thus it
has been used to reconstruct interaction patterns of complex systems, such as the coupling constants in a magnetic alloy, 
the interactions between firing neurons in neural networks (either in vivo, in vitro or in silico), the way chemical reaction are coupled together in metabolic networks, sociological interactions in social network, etc. 
The applications of the inverse Ising model can be found in 
in different fields of science including
physics \cite{cocco2011adaptive,ricci2011mean,inf_nguyen-11,inf_nguyen-12,aurell2012inverse,zhang12,decelle13}, computer science \cite{wainwright2010AnnalHigh},  neuroscience
 \cite{Schneidman_etal_2006_nature, roudi2009}, social network \cite{fortunato2010community} and biology \cite{ruz2010learning,weigtPNAS,ekeberg2013improved,bialek2012statistical}. 

The canonical approach to tackle this problem is by the inference methods that 
maximize the likelihood of the parameters given the data.
Most of the studies on this subject focused on improving the performance of 
the inference by increasing accuracy and efficiency of inference methods \cite{cocco2011adaptive,ricci2011mean,inf_nguyen-11,inf_nguyen-12,aurell2012inverse,zhang12,decelle13,decelle15,ferrari2015approximated}, 
for instance by applying improved mean-field and
cluster variational methods \cite{raymond2013mean}, by using regularizations of different forms \cite{wainwright2010AnnalHigh}, etc. 
A similar and very effective approach can be used also in case data comes from a dynamical process, i.e.\ the so-called kinetic Ising model \cite{roudi2011mean,mezard2011exact,zeng2011network,zhang12,zeng2013maximum,decelle15,bachschmid2015learning}, although most of these studies focus on the simpler asymmetric model where $J_{ij}$ and $J_{ji}$ are independent couplings.

However little attention has been paid to the data side, for instance how to 
improve the performance of inference by increasing the quality of data.
We think that, in the contemporary ``age of big-data'', the data-side consideration 
could become more and more important, as recently we have been observing that the amount of available data in many fields has been growing so quickly that, in some cases, taking all of them into account for the inference becomes a computationally difficult task.
Then a natural question arises: do we really need all the data for the inference? 
To put it differently, does every configuration in the dataset contain equal amount of information about the system?
Obviously, this question concerns the \textit{data quality} problem: how to
quantify the quality of the data and how to eventually improve it.

In this article we address directly the questions posed in the last paragraph -- the data quality in the inverse Ising problems. 
We will focus in this work on the dynamical inverse Ising case where the data are generated by a stochastic process where variables get updated synchronously. Using this model, we will demonstrate that data coming from the out-of-equilibrium regime are much more informative than equilibrium configurations, or configurations gathered from a steady state of the system.
Our results may suggest new experimental protocols to acquire data used to reconstruct the interaction network: the system under study should be first perturbed to an out-of-equilibrium state, and then measured.

The paper is organized as follows.
Sec.~\ref{ref_model} contains descriptions of the model and the stochastic process to update/evolve the system variables. 
In Sec.~\ref{ref_dataq} we quantify the
data quality using the effective rank of the correlation matrix, and show that out-of-equilibrium data has higher quality than the equilibrium/stationary data.
In Sec.~\ref{sec:decimation} we propose a method based on the perturbation analysis of the correlation matrix for tuning the quality of data, i.e. removing configuration from the data in such a way that the data quality keeps improving.
Finally we conclude this work in Sec.~\ref{sec:con}.

\section{The dynamical inverse Ising model}
\label{ref_model}

The dynamical inverse Ising model is based on pairwise interactions amongst discrete variables and a dynamical rule to update these variables. 
The general setting considers $n$ variables corresponding to the $n$ nodes of a graph. Each variable (denoted by $i$) takes values $s_i = \pm 1$. 
An edge or coupling between node $i$ and node $j$, $J_{ij}$, takes a real value to represent an interaction between two nodes. 
In this work we consider only the case of symmetric interactions, i.e.\ $J_{ji} = J_{ij}$, which ensures the existence of an equilibrium (i.e.\ stationary) measure $P_{\rm eq}(\us)$ at inverse temperature $\beta$ (the generalization for non-symmetric couplings is straightforward)\footnote{As usual, we scale the interactions such as to have unitary variance (in the units that make the energy extensive) and avoid self-interactions ($J_{ii}=0$).}.

A common and practical choice for the stochastic process that simulates the system evolution is 
the so-called parallel dynamics, where a new configuration at time $t+1$ is 
drawn synchronously from the state at time $t$:
at each time step, each variable is updated according to
the local field acting on it, defined as $h_i^{\rm loc}(\underline{s}) = \sum_{j} J_{ij} s_j$.
Then, the probability of the configuration at time $t+1$ can be written as
\begin{equation}
    P[\underline{s}^{(t+1)}|\underline{s}^{(t)}] = \prod_{i=1}^n \frac{e^{\beta s_i^{(t+1)} h_i^{\rm loc}(\underline{s}^{(t)})}}{2 \cosh(\beta h_i^{\rm loc}(\underline{s}^{(t)}))}\,.
    \label{eq:dyn}
\end{equation}
It can be shown that in general this dynamics respects the detailed-balance (some oscillations can arise for $\beta \to \infty$, but it is not of our concern here). However, it is known that, in many cases, the dynamics can be very slow to reach thermal equilibrium. It is typically the case when the system is in a glassy phase (e.g.\ a Sherrington-Kirkpatrick model \cite{sherrington1975solvable} with $\beta>1$), or when it undergoes a rapid quench beyond a second order phase transition. 

Let us consider the following experiment. First, we generate the dataset: for a given set of couplings $\underline{J}\equiv\{J_{ij}\}$ and a random uniform initial condition $\underline{s}^{(0)}\in\{-1,1\}^n$, we generate $L$ correlated configurations $\underline{s}^{(t)}$ for $t=1,\ldots,L$, using the parallel dynamics in Eq.~(\ref{eq:dyn}).
Second, we try to infer the values of the couplings $\underline{J}$ using only the data $\{\underline{s}^{(t)}\}_{t=0,\ldots,L}$. Since the dynamical updating rule is known, we can achieve this goal by maximizing the empirical log-likelihood,
\begin{align}
    \mathcal{L} & = \langle\log( \prod_{t=1}^L P[\underline{s}^{(t)}|\underline{s}^{(t-1)}] )\rangle_\data \nonumber \\
     & = \langle \sum_{t=1}^L \sum_{i=1}^n \beta s_i^{(t)} h_i^{\rm loc}(\underline{s}^{(t-1)}) - \log( 2 \cosh(\beta h_i^{\rm loc}(\underline{s}^{(t-1)})) )\rangle_\data \nonumber\\
     & = \sum_{t=1}^L \bigg[ \sum_{i,j} \beta J_{ij} \langle s_i^{(t)} s_j^{(t-1)} \rangle_\data - \sum_i \langle \log( 2 \cosh(\beta h_i^{\rm loc}(\underline{s}^{(t-1)})) ) \rangle_\data\bigg],
	\label{eq:likelihood}
\end{align}
where $\langle\cdot\rangle_\data$ denotes the average over the configurations generated.
At variance to the static inverse Ising problem, we are able 
to maximize directly the likelihood for this dynamical model, since all the terms in
Eq.~(\ref{eq:likelihood}) can be computed in polynomial time.

To evaluate the performance of reconstruction we consider a measure 
to the difference between inferred couplings $\underline{J}$ and 
the true couplings $\underline{J}^*$:
\begin{equation}\label{eq:diff}
	\Delta_J = \sqrt{\frac{\sum_{i<j} \left(J_{ij}-J_{ij}^* \right)^2 }{n(n-1)/2}}\,.
\end{equation}
In cases where the system under study is an instance of ensemble of problems, i.e.\ it is a disordered model, we should also average the reconstruction error over the disorder ensemble.
However, we expect the reconstruction error to be self-averaging, so few samples are enough to estimate it.
In practice, for each value of the parameters we are going to use we choose a different sample (i.e.\ different couplings), such that sample-to-sample fluctuations can be appreciated in the plots reporting the results on $\Delta_J$.

\section{Data quality}
\label{ref_dataq}

In many situations we have the freedom to decide how the data is acquired in the experiments and in the real-world inference problems. This may give a way to select high-quality data rather than poor-quality one.
In this paper, we consider an experimental setting where we are able to gather
configurations either in-equilibrium or out-of-equilibrium.
The details of the protocol are described below:
\begin{enumerate}
\item First, $m$ initial configurations are randomly and uniformly chosen in $\{-1,1\}^n$.
\item For each initial configuration, $T$ steps of parallel dynamics are performed, and the final configuration is recorded. This
process generates $m$ configurations that we store in the rows of the matrix
$A\in \{-1,1\}^{m\times n}$.
\item For each configuration stored in $A$, we do a single step of parallel dynamics, and then record the new configurations as the rows of the matrix $B$.
\item Lastly, we infer the couplings using matrices $A$ and $B$.
\end{enumerate}
The likelihood is proportional to the probability of generating matrices $A$ and $B$ given couplings $\underline{J}$
\[
P(A,B;\underline{J}) = P(B|A;\underline{J}) P(A;\underline{J})
\]
We chose to infer the most likely couplings $\underline{J}$ by maximizing $P(B|A;\underline{J})$ with respect to $\underline{J}$.
In principle some information is also contained in the term $P(A;\underline{J})$, but we ignore such an information for the following reasons. Maximizing $P(A;\underline{J})$ is computationally very demanding: in the equilibrium/stationary limit ($T \gg 1$) it corresponds to solving the ``static'' inverse Ising problem, while for small values of $T$ the amount of information in $P(A;\underline{J})$ is limited (it is null for $T=0$) and hard to extract given that $P(A;\underline{J})$ is not a Boltzmann-Gibbs like distribution.

The explicit expression of the log-likelihood, $\log P(B|A;\underline{J})$, in terms of matrices $A$ and $B$ is
\begin{equation}
\mathcal{L}  = \sum_{a=1}^{m}\brb{\sum_{i,j} \beta J_{ij} B_{ai}A_{aj} - \sum_i \log( 2 \cosh(\beta \sum_j J_{ij} A_{aj}) )}.
	\label{eq:likelihood2}
\end{equation}
At the stationary point of $\mathcal{L}$, the derivative with respect to each coupling $J_{ij}$ must be zero and this leads to the following moment-matching condition
\begin{equation}
\label{eq:momMatch}
\langle A_{ai}\,B_{aj} \rangle_\data = \langle A_{ai}\tanh(\beta \sum_k J_{jk} A_{ak})\rangle_\data\,,
\end{equation}
where the average over the data is given by $\langle \bullet \rangle_\data = m^{-1}\sum_{a=1}^m \bullet$, and the following updating rule for the couplings
\begin{equation*}
J_{ij} \leftarrow J_{ij} + \eta \left[ \langle A_{ai}\,B_{aj} \rangle_\data - \langle A_{ai}\tanh(\beta \sum_k J_{jk} A_{ak})\rangle_\data\right]\,,
\end{equation*}
being $\eta$ a small learning parameter.

With respect to the standard inverse kinetic Ising problem, our experimental setup has two main differences. (i) The new parameter $T$ allows us to collect configurations both in the stationary equilibrium regime (as usual) for a large $T$, but also in the early out-of-equilibrium regime, for small $T$, where configurations are sampled according to a probability distribution different from $P_{\rm eq}(\us)$. (ii) For each initial configuration we do not save the entire trajectory but only the last two configurations, corresponding to times $T$ and $T+1$. This choice allows us to better understand how the data quality depends on the ``distance from equilibrium''. Moreover a longer trajectory of $L$ steps can always be seen as the union of $L-1$ of our one-step experiments, with different $T$ parameters (this will be further discussed in Section \ref{sec:multi}).

The reasons why we expect out-of-equilibrium configurations to be of higher quality for the problem of coupling reconstruction are possibly many.
First of all, since we start from $m$ random configurations, we have that the configurations in matrix $A$, measured at time $T$, are less correlated and spanning a broader region of the configuration space, with respect to equilibrium configurations; in general, we expect correlations between configurations to increase monotonously with $T$. Moreover in the early out-of-equilibrium regime the dynamics usually has some drift, which is absent at equilibrium: e.g.\ the energy decreases towards the equilibrium value, and then stays more or less constant. Generally in the out-of-equilibrium early dynamics the system variables get updated more often, and this may lead to a sensible increase in the measured correlations and fluctuations, which are in turn exploited by the moment matching condition in Eq.~(\ref{eq:momMatch}) to infer the couplings.

We are mostly interested in studying what happens in the low temperature regime (large $\beta$) because in the low $\beta$ regime correlations are weak, and many efficient methods exist for inferring the couplings. The low temperature dynamics, starting from a random initial condition, strongly depends on the kind of system under study (whether it is homogeneous, heterogeneous, with disordered couplings, etc.); however in general the dynamics shows an initial fast relaxation, when most of the variables get updated often. On later times, the dynamics can easily get trapped in a local energy minimum, keeping oscillating around it: it is clear that this asymptotic regime is much less informative for coupling reconstruction, because only a small fraction of variables keep updating and usually in a repetitive way.

Our goal here is not to characterize in detail the behavior of the inference algorithm for different values of $T$ and $m$. We are rather interested in showing that, intrinsically, out-of-equilibrium configurations contain more information than equilibrium ones. In order to show that, we will use some particular values of $T$ and $m$. As said earlier, we will focus only on the two last configurations at time $T$ and $T+1$. We shall then fix the value of $m$ to a particular value and look at different values of $T$ to probe both the out-of-equilibrium regime, for small $T$ values, and the equilibrium or steady-state regime, for large $T$ values.

In order to study an interesting and difficult case we consider the Sherrington-Kirkpatrick (SK) model \cite{sherrington1975solvable} where couplings $\{J_{ij}\}_{i<j}$ are randomly and independently extracted from a Gaussian distribution of zero mean and variance $1/n$ (while for $i>j$ we set $J_{ij}=J_{ji}$). We will consider small systems ($n=20$) because our results are of a general validity and do not require to take the thermodynamical limit ($n\gg 1$).

\begin{figure}
   \centering
   \includegraphics[width=0.45\linewidth]{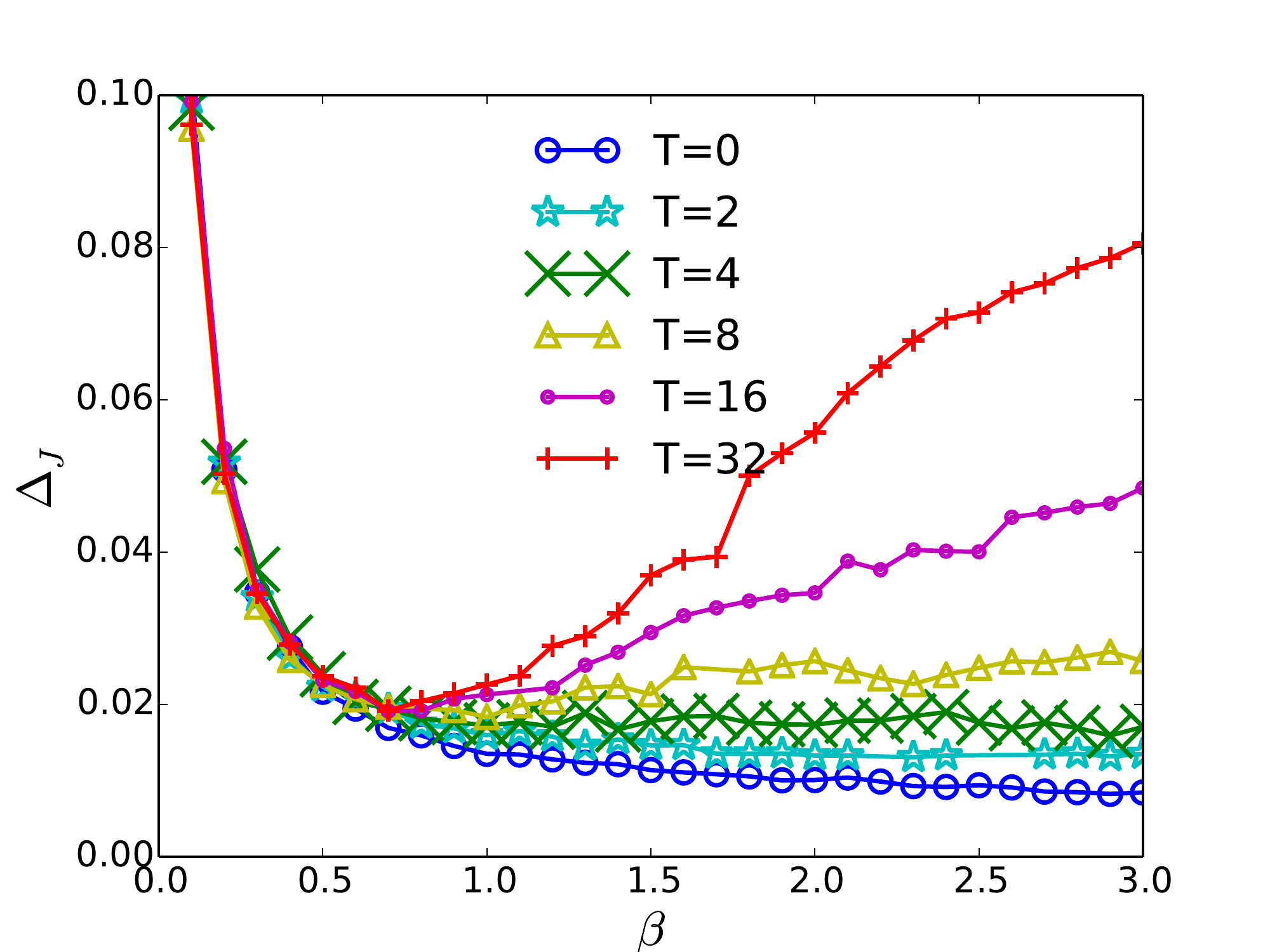}
   \includegraphics[width=0.45\linewidth]{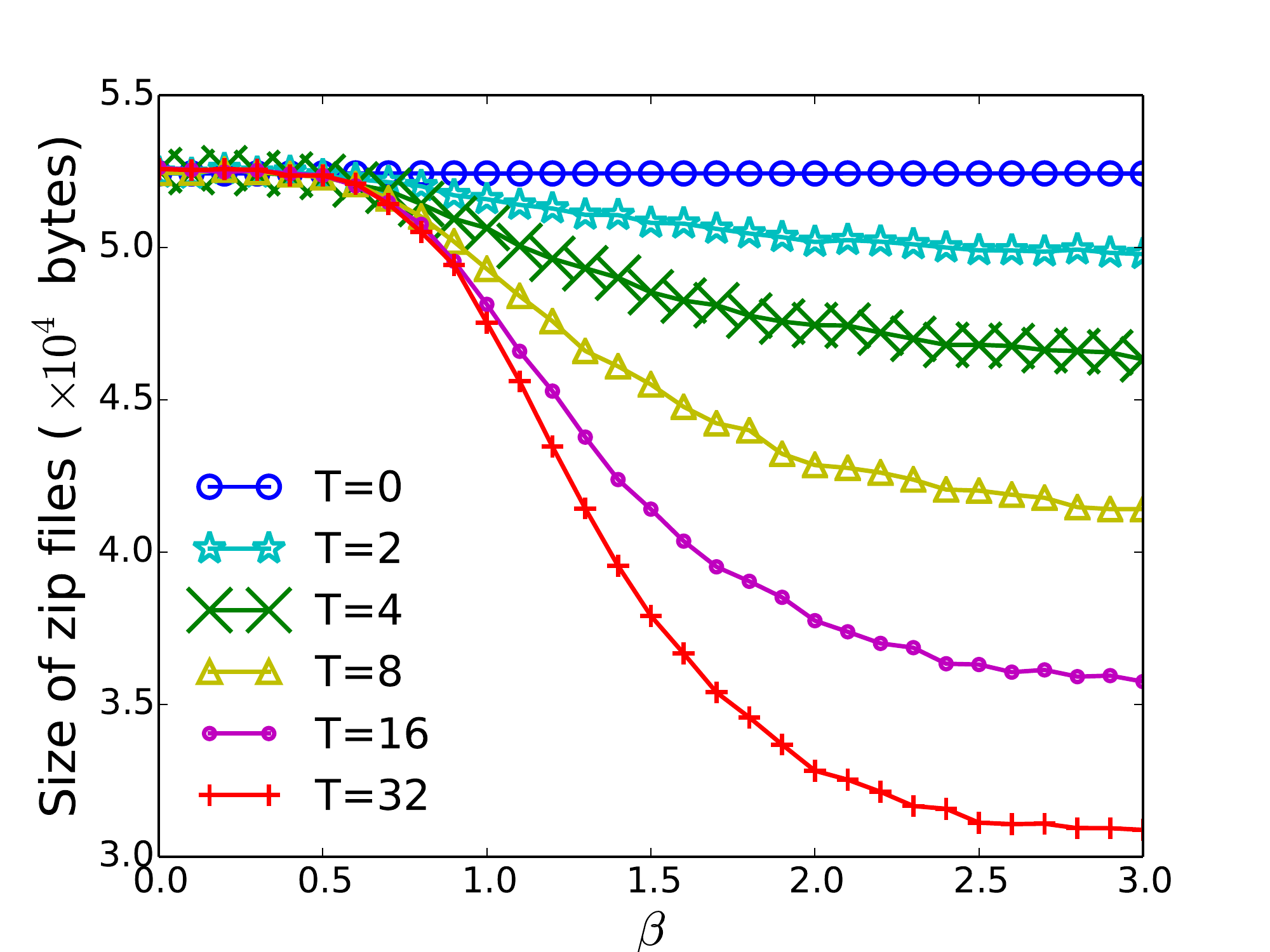}
   \includegraphics[width=0.48\columnwidth]{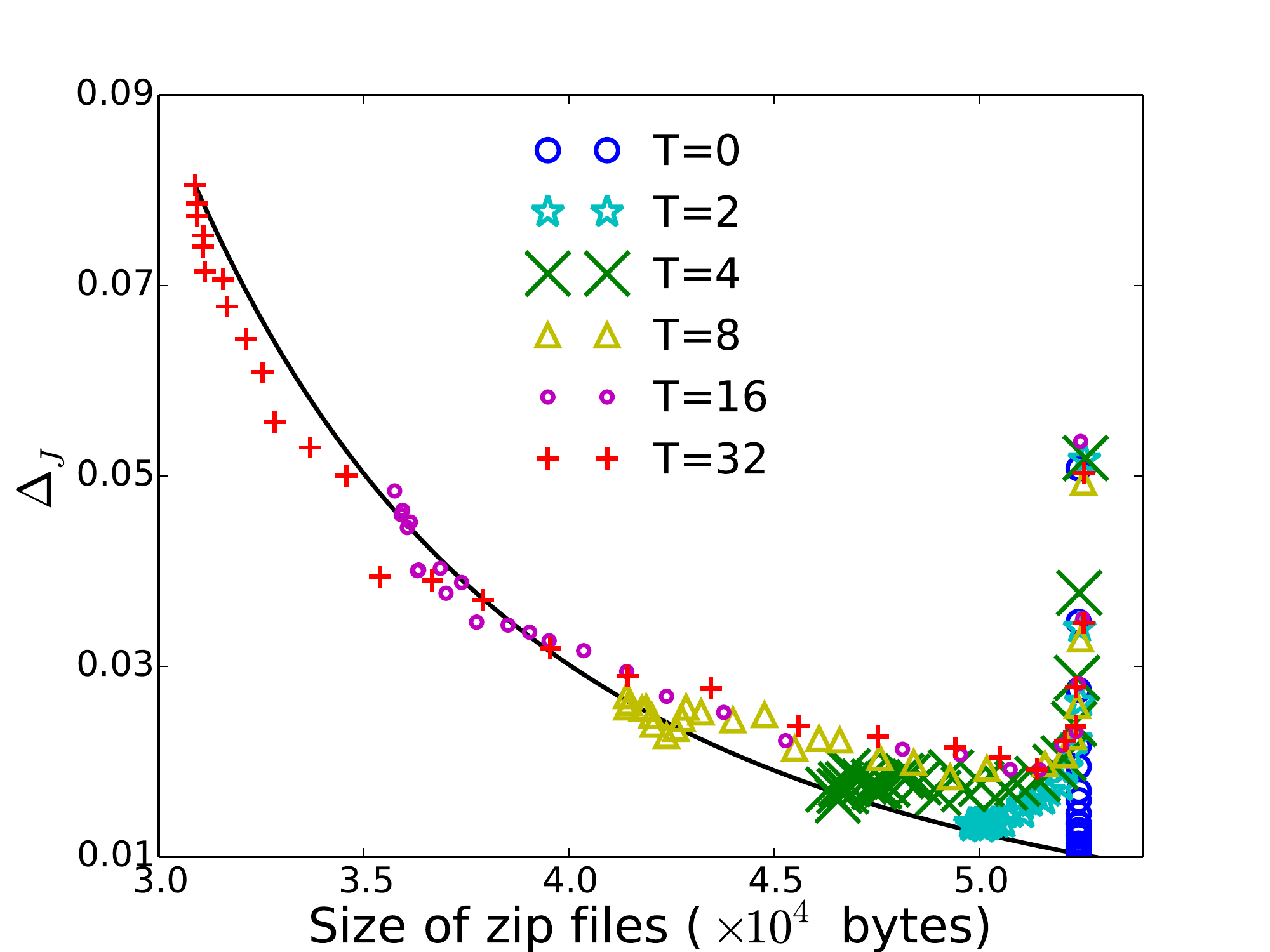}
   \caption{Inference error (\textit{upper left}) and size of zipped files (\textit{upper right}) for several $T$ and varying $\beta$. \textit{Lower panel}: relation between the reconstruction error and the size of zipped files, with the black line being a power law fit with exponent $3.7$.
	Experiments were carried out on a network with $n=20$ spins, using $m=10^4$ configurations.
   \label{fig:entropy}}
\end{figure}

In Figure \ref{fig:entropy} we show the results of experiments performed with $m=10^4$ random restarts and several values of $T$. For each value of $T$ and $\beta$ we use a different sample (i.e.\ couplings), such that fluctuations in the data points are meaningful for estimating the sample-to-sample variations.
In the upper left panel we show how the reconstruction error varies with $\beta$ for several different values of $T$. We see basically two regimes where the reconstruction error behave differently. In the regime $\beta \lesssim 1$ the reconstruction error is roughly the same for any $T$ value. This regime corresponds to the paramagnetic phase of the SK model, where ergodicity ensures that configurations remain mostly uncorrelated for any value of $T$, even approaching equilibrium. The regime $\beta \gtrsim 1$ corresponds to the glassy phase of the SK model, where ergodicity is broken. In this regime the out-of-equilibrium configurations (gathered at small $T$ values) provide a clearly better quality for the inverse Ising problem, resulting in a much smaller reconstruction error. The figure also shows that the reconstruction error grows monotonically with $T$, thus becoming larger and larger when the dynamics brings the system close to equilibrium.

On a first sight, this result may appear counterintuitive: increasing $T$ there is an entropy decrease in the model and consequently one would expect to have a gain of information; instead we observe an increasing error. We stress once more that this gain of information would be observed in the likelihood $P(A;\underline{J})$, that we decided not to use (for the reasons already discussed above). For the likelihood $P(B|A;\underline{J})$ things go the other way around: as we already discussed above, there may be several reasons for the increase of the reconstruction error when configurations are sampled closer to equilibrium. Among these, one possibility is that the $m$ configurations sampled for $\beta>1$ and $T>0$ are somehow similar and thus redundant.
In practice the $m \times n$ matrix $A$ has correlated entries and we would like to measure how much one can reduce it without losing information.
The simplest way to achieve this is to run an efficient algorithm for lossless compression: we use \textit{gzip} to compress each $A$ matrix and we measure the size of the compressed file.
In upper right panel of Fig.~\ref{fig:entropy}, we plot the size of the compressed files
for several values of $T$ and varying $\beta$. A comparison with the curves in upper left panel of the same figure suggests that the increase of the reconstruction error in the low temperature phase ($\beta \gtrsim 1$) is mostly related to the loss of information in the $m$ configurations used to infer the couplings.
On the contrary the increase of the error in the high temperature limit ($\beta \to 0$) is due to the lack of correlations among variables, that makes impossible to extract information about the couplings.

Let us call $N_\text{eff}$ the effective size of the $A$ matrix as measured from the size of the compressed file (the $B$ matrix has approximately the same effective size). A naive expectation would be that the reconstruction error grows like $1/\sqrt{N_\text{eff}}$, when the effective size of the $A$ matrix decreases. However the lower panel in Fig.~\ref{fig:entropy} shows that the increase in the reconstruction error by decreasing $N_\text{eff}$ is much steeper: the power law curve shown in the figure has slope $3.7$, thus suggesting that the error increase also depends on other factors, like (i) the fact that equilibrium configurations evolve more slowly, and thus the number of spin flips in a one-time experiment is smaller and (ii) the presence of long-ranged spatial correlations, that grow approaching the equilibrium.

\begin{figure}
   \centering
   \includegraphics[width=0.58\columnwidth]{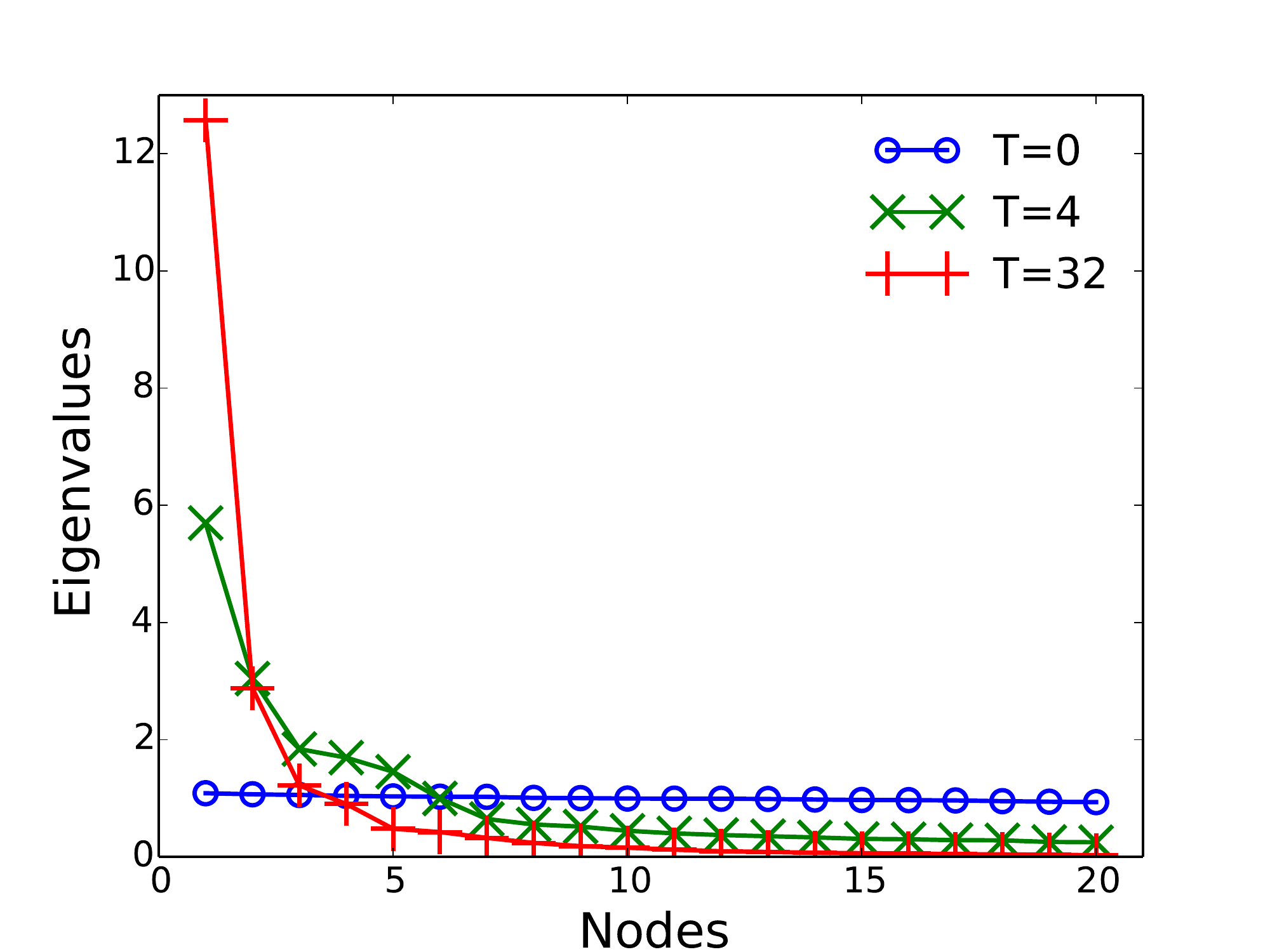}
   \includegraphics[width=0.48\columnwidth]{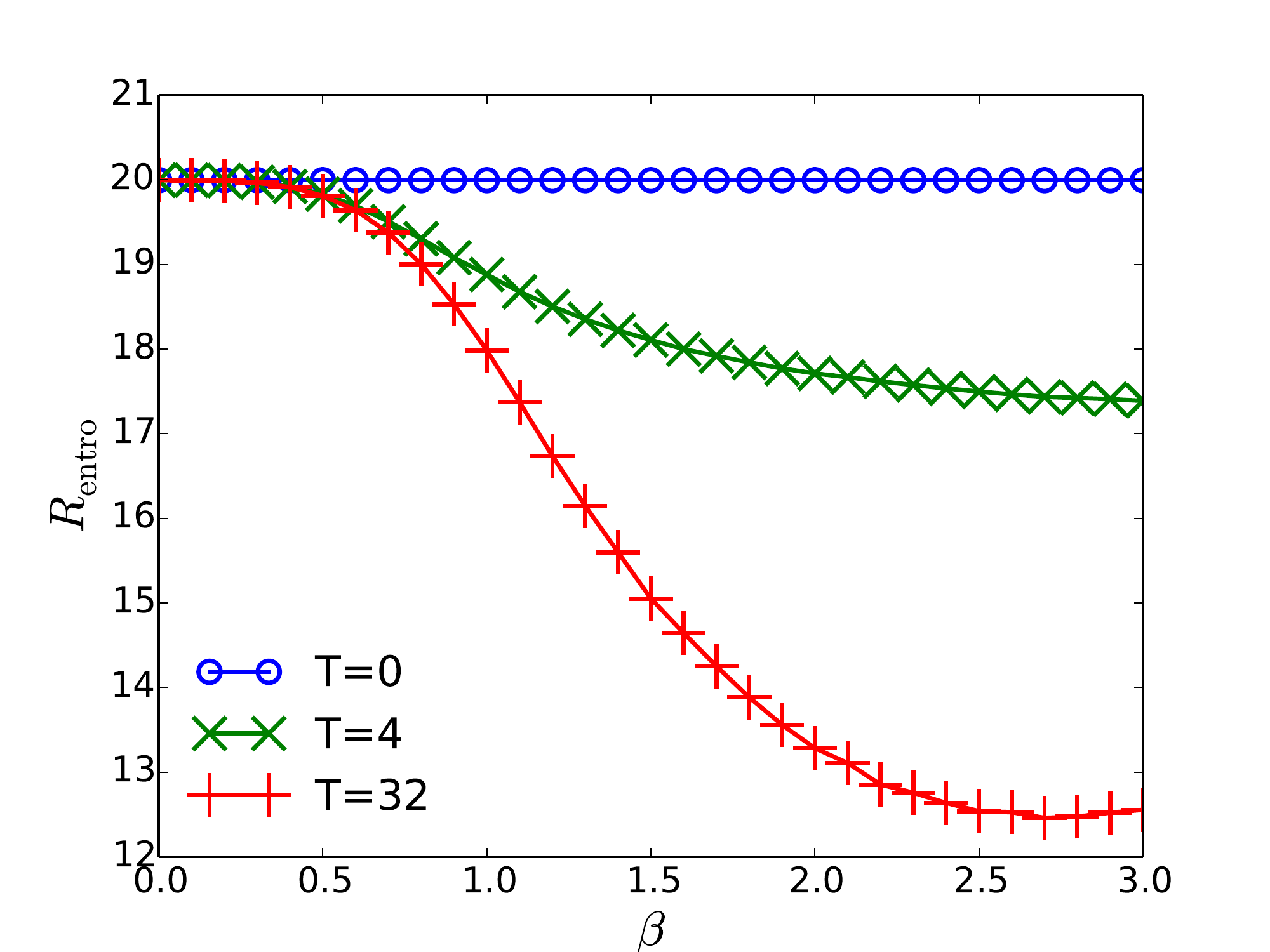}
   \includegraphics[width=0.48\columnwidth]{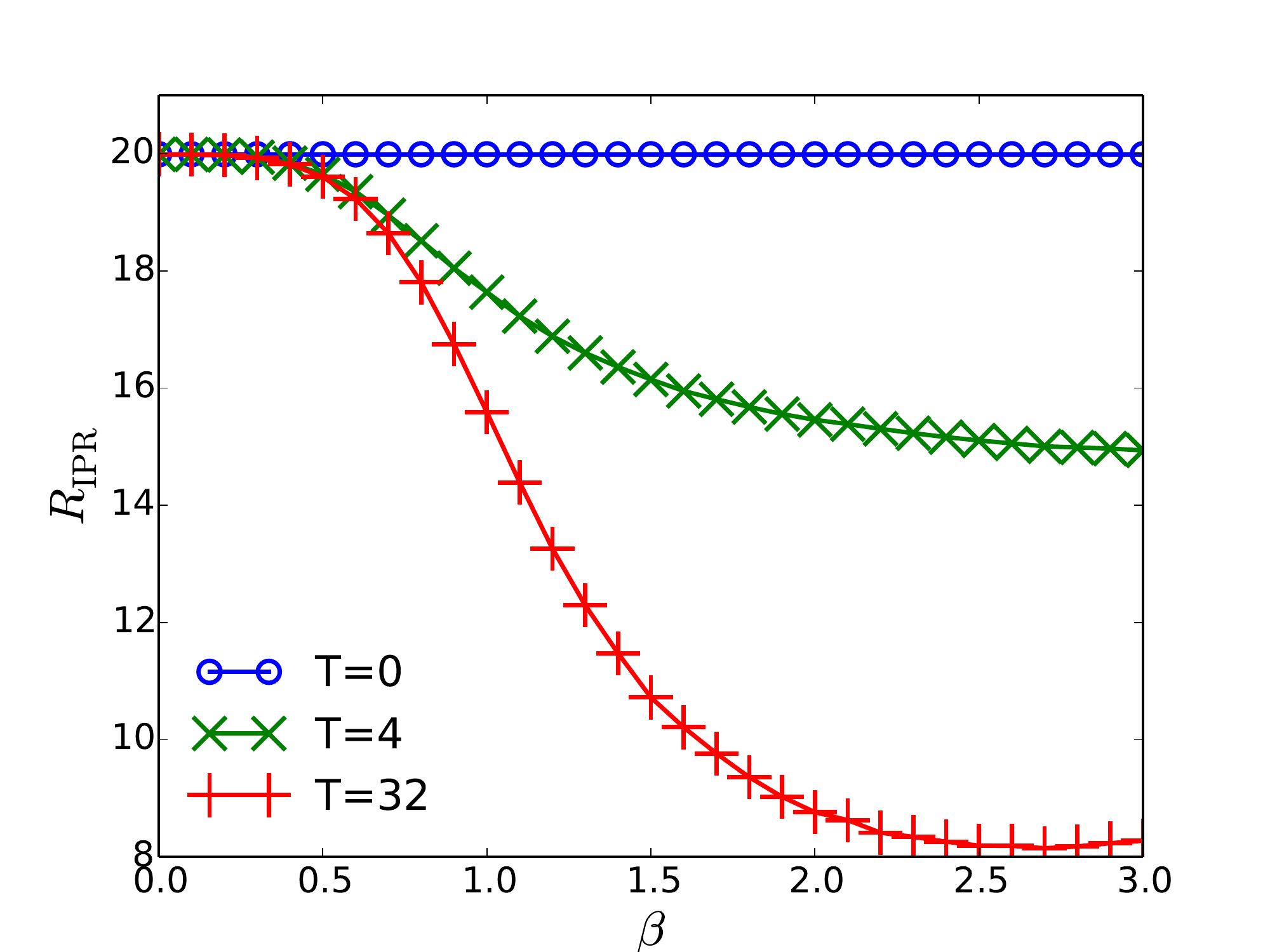}
   \caption{\textit{Top panel}:
   Eigenvalues of the correlation matrix for $T=0,4,32$ at $\beta=2.2$.
   Effective rank $R_\text{entro}$ (\textit{left panel}) and $R_\text{IPR}$ (\textit{right panel})  of the correlation matrix for $T=0,4,32$ and varying $\beta$.
Parameters are the same as in Fig.~\ref{fig:entropy}.
   \label{fig:pca}}
\end{figure}

A more formal way to define the effective size of the matrix $A$ is to compute its \emph{effective rank}, i.e.\ a measure of how much correlated are the entries of the matrix $A$.
In principle we would like to do the principal component analysis (PCA) of the configuration matrix $A$, which would tell us whether there are preferred directions along which configurations tend to align.
In practice we consider the eigen-decomposition of the correlation matrix $C$, which is defined as
\begin{equation*}
	C=\frac{1}{m} A^T A\,.
\end{equation*}
In our experiments we have $2^n\gg m \gg n$, thus $C$ has $n$ real eigenvalues 
$\{\lambda_i\}_{i\le n}$, satisfying
\begin{equation}
	\label{eq:sum}
	\sum_{i=1}^n\lambda_{i}=\sum_{i=1}^nC_{ii}=n\,.
\end{equation}
In the top panel of Fig.~\ref{fig:pca} these eigenvalues are plotted in a decreasing order for some $T$ values. We can see that for $T=0$, when configurations are completely random, the distribution of eigenvalues is flat and every eigenvalue is close to $1$.
It means that there is no particularly preferred direction and the vectors in $A$ span uniformly the configurational space, thus providing $m$ configurations with practically zero redundancy.
However with $T=32$, being configurations closer to equilibrium, they tend to align along a preferred direction and the first eigenvalue, corresponding to this direction, is much larger than the other eigenvalues.
Hence, we see that the distribution of the eigenvalues can be used to characterize the data quality of configurations. 

From Eq.~\eqref{eq:sum} we know that the sum of 
all the eigenvalues is a constant for different sets of configurations of the same system.
So the most naive way to measure how flat is the distribution of eigenvalues,
is to compute the largest eigenvalue (in absolute value):
the smaller the leading eigenvalue is, the flatter the distribution of the eigenvalues and therefore better the quality of the data.
A more comprehensive approach is to compute the effective rank of the matrix $C$, defined by \cite{roy2007effective}
\benn
\text{rank}_\text{eff}(C) = \exp\Big(-\sum_{i=1}^n p_i \log p_i\Big)\,,
\eenn
where $p_k = \lambda_k / \sum_{i=1}^n \lambda_i = \lambda_k / n$.
Again, an almost flat distribution of eigenvalues ($p_i\simeq 1/n$) implies the rank of $C$ is close to $n$, while the effective rank decreases if the eigenvalues are very different among them.

Actually we are mostly interested in understanding how much redundant are the vectors by which the matrix of empirical correlations C is build, rather than the matrix itself. Being the eigenvalues of $C$ real and positive, we can write $\lambda_i=\sigma_i^2$ and the following decomposition
\be
\label{eq:Cdecomp}
C = \sum_{i=1}^n \lambda_i\,\uv_i\,\uv_i^T = \sum_{i=1}^n (\sigma_i \uv_i) (\sigma_i \uv_i)^T\,,
\ee
where $\{\uv_i\}$ are the eigenvectors of $C$, forming an orthonormal basis. Eq.~\eqref{eq:Cdecomp} says that the same $C$ matrix could be obtained if the measured configurations were only equal to one of the eigenvectors $\{\uv_i\}_{i\le n}$, each one chosen with a probability $r_i = \sigma_i / \sum_{k=1}^n \sigma_k$.
From these frequencies $\{r_i\}_{i\le n}$ we can provide two different, but similar, definitions of the effective rank of matrix $C$
\bea
R_\text{entro} &=& \exp(-\sum_{i=1}^n r_i \log r_i)\,,\nonumber\\
R_\text{IPR} &=& \frac{1}{\sum_{i=1}^n r_i^2}\,.\label{eq:rank}
\eea
The first definition is simply based on the entropy of the probability law $\{r_i\}_{i\le n}$, while the second definition is the inverse participation ratio. Both effective ranks $R_\text{entro}$ and $R_\text{IPR}$ take values in $[1,n]$: they are equal to $n$ if $r_i=1/n$, and equal to 1 if the probability concentrates on a single value.

In the lower panels of Fig.~\ref{fig:pca} we plot the effective ranks $R_{\rm{entro}}$ and $R_{\rm{IPR}}$ for the same data used in 
Fig.~\ref{fig:entropy}. We can see that they give similar information as the size of the compressed file:
the smaller the effective rank, the worse data quality. The advantage of using 
the effective rank $R$ over using the size of the compressed file is that the effective rank is 
easier to compute. We can thus use it as an objective function to optimize the data quality, as we will show in the next section.

\section{Tuning quality of the data}
\label{sec:decimation}

In this section we study how to identify configurations that have
relatively bad quality in a given set of configurations.
That is those configurations that, being redundant, can be safely removed from the set without loosing too much information, and thus actually improving the data quality.
The idea is that if we remove a configuration from the dataset, all 
eigenvalues of the correlation matrix $C$ will shift from
$\bre{\lambda_i}$ to $\bre{\lambda_i+\Delta {\lambda_i}}$.
Thus we can estimate the quality of each configuration in the dataset, according to the shift of 
the effective rank $R$ in case that configuration is removed.
We aim at removing configurations in the direction of increasing $R$, in order 
to improve the data quality of set of remaining of configurations.

Since the number of configurations is large, we can treat the effect of removing
one configuration, i.e. one row in matrix $A$, as a perturbation to the correlation 
matrix $C$. That is, after removing the configuration $\us$, the change of 
$C$ is
$$\Delta C=\frac{C-\us\,\us^T}{m-1}.$$
Assuming that after removing the configuration $\us$, the $i$-th eigenvector of $C$ changes from 
$\uv_i$ to $\uv_i+\Delta \uv_i$, and its associated eigenvalue changes from $\lambda_i$ to 
$\lambda_i+\Delta\lambda_i$, then we have 
$$(C+\Delta C)(\uv_i+\Delta \uv_i)=(\lambda_i+\Delta\lambda_i)(\uv_i+\Delta \uv_i).$$
Keeping only first-order terms results in
\begin{equation*}
	\Delta\lambda_i=\uv_i^T\Delta C\,\uv_i.
\end{equation*}
Then by making use of Eq.~\eqref{eq:rank} and by keeping only the first order of 
$\Delta {\lambda_i}$, we can estimate the shift of effective rank as
\begin{equation}
	\label{eq:deltaR}
	\Delta R_\text{IPR}=\sqrt{\frac{R_\text{IPR}}{n}} \sum_i \lambda_i^{-\frac{1}{2}}\Delta \lambda_i
\end{equation}

Then, using Eq.~(\ref{eq:deltaR}), we propose a decimation method to increase the data quality by removing iteratively configurations that provide the largest $\Delta R$.
This procedure is similar to the decimation algorithm using marginals of a message passing algorithm in solving constraint satisfaction problems \cite{Mezard2002}, where nodes having most biased marginals are removed (fixed) at each iteration.

\begin{figure}
   \centering
   \includegraphics[width=0.48\columnwidth]{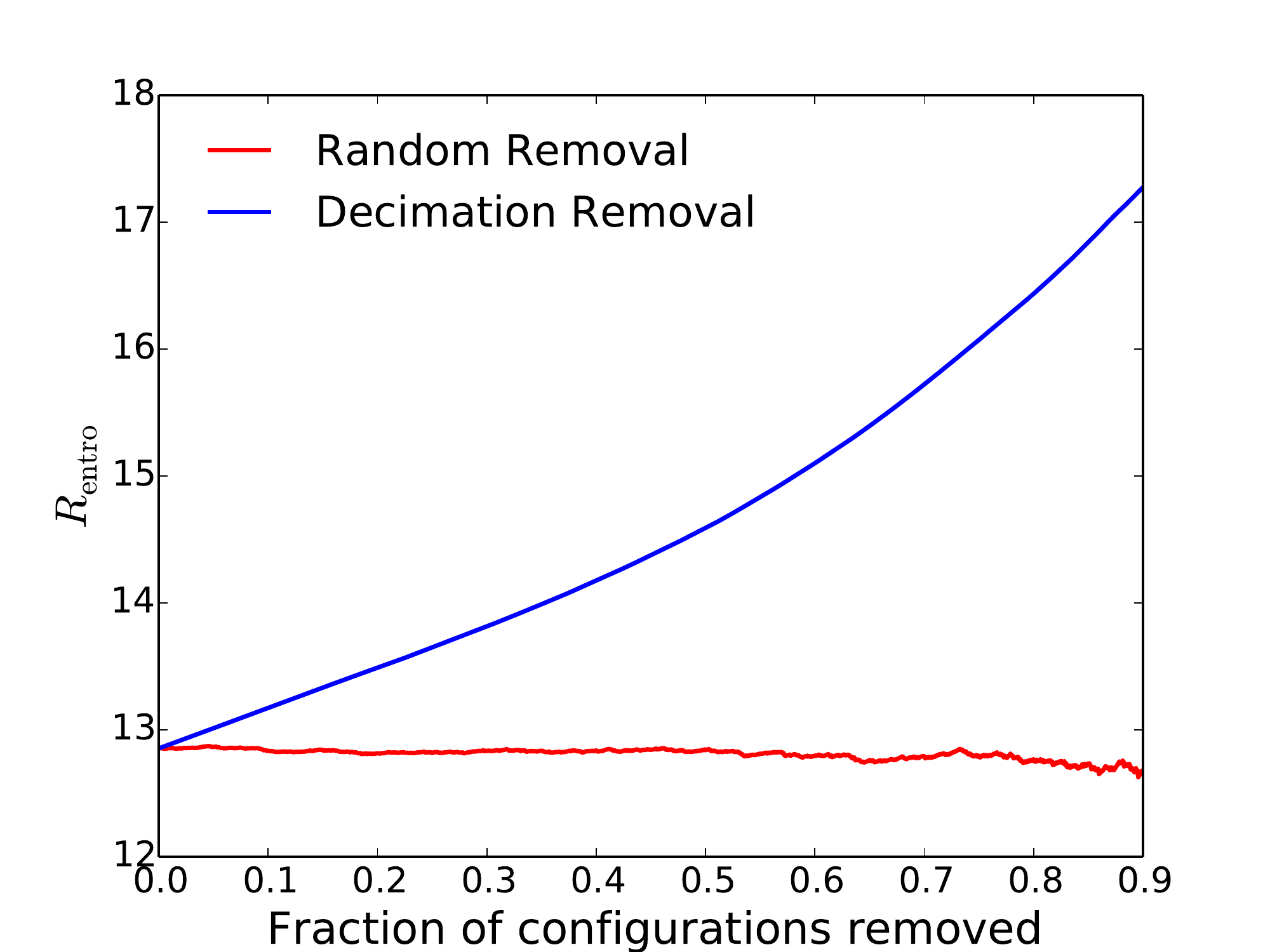}
   \includegraphics[width=0.48\columnwidth]{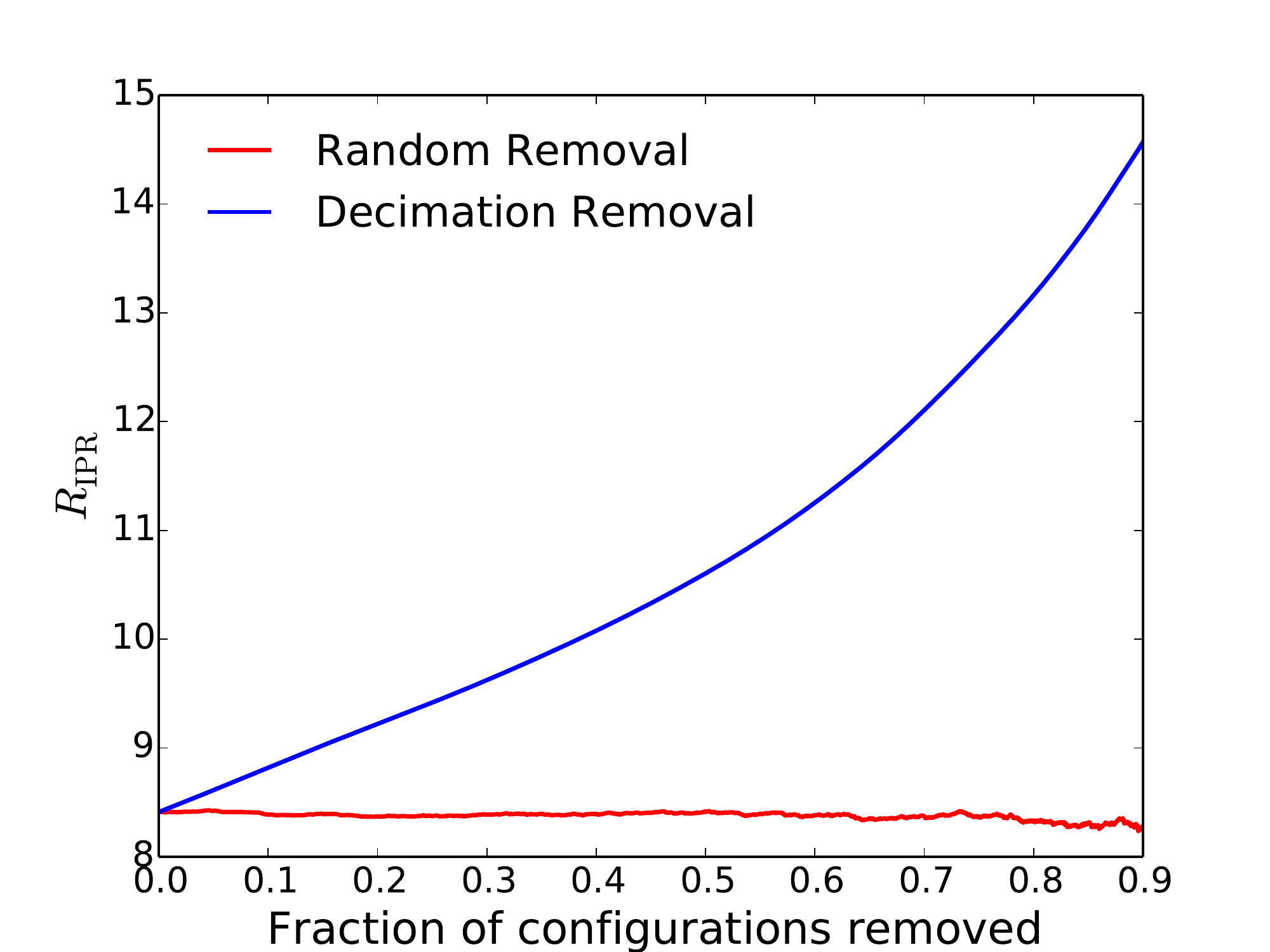}
   \includegraphics[width=0.48\columnwidth]{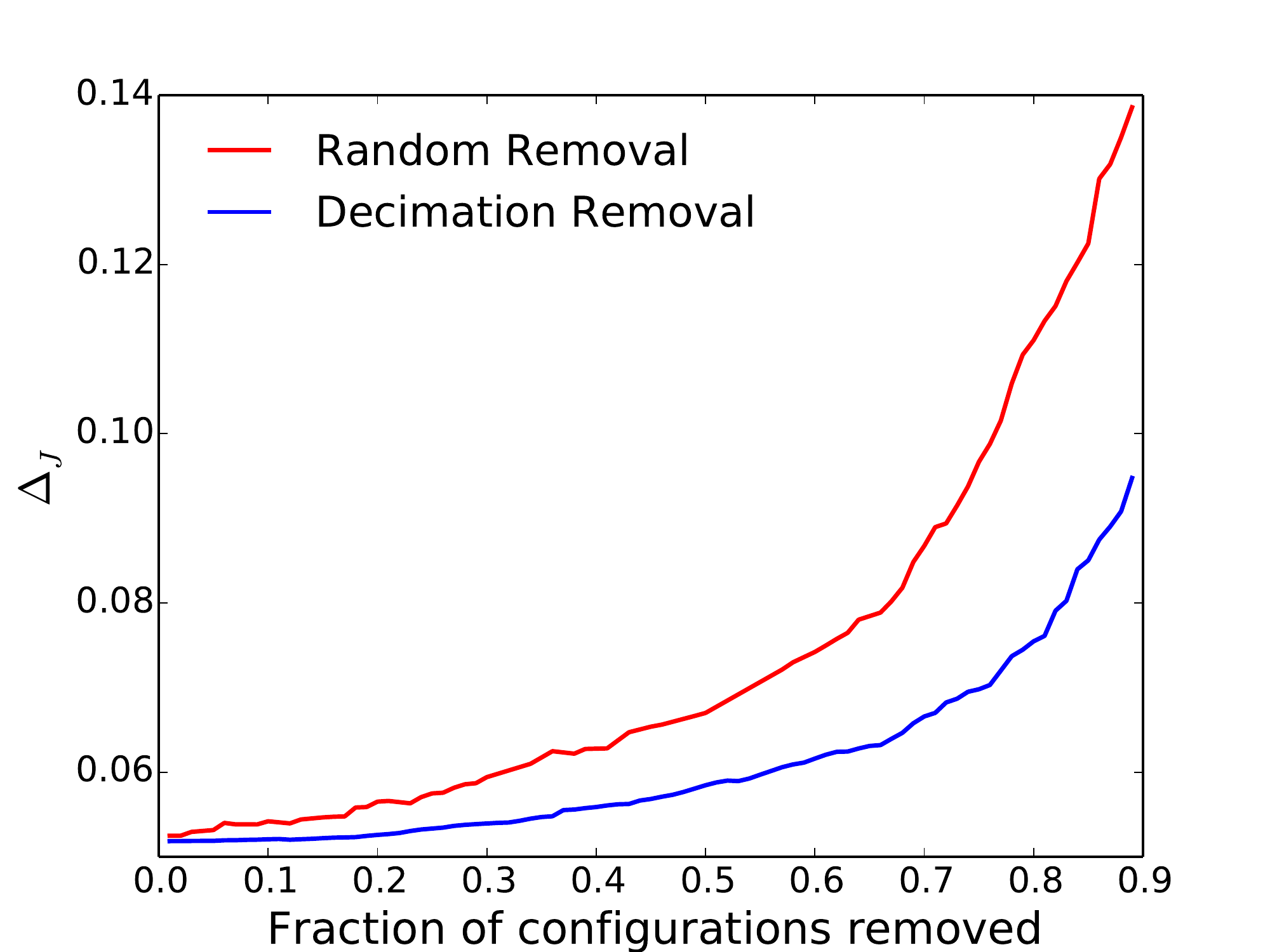}
   \includegraphics[width=0.48\columnwidth]{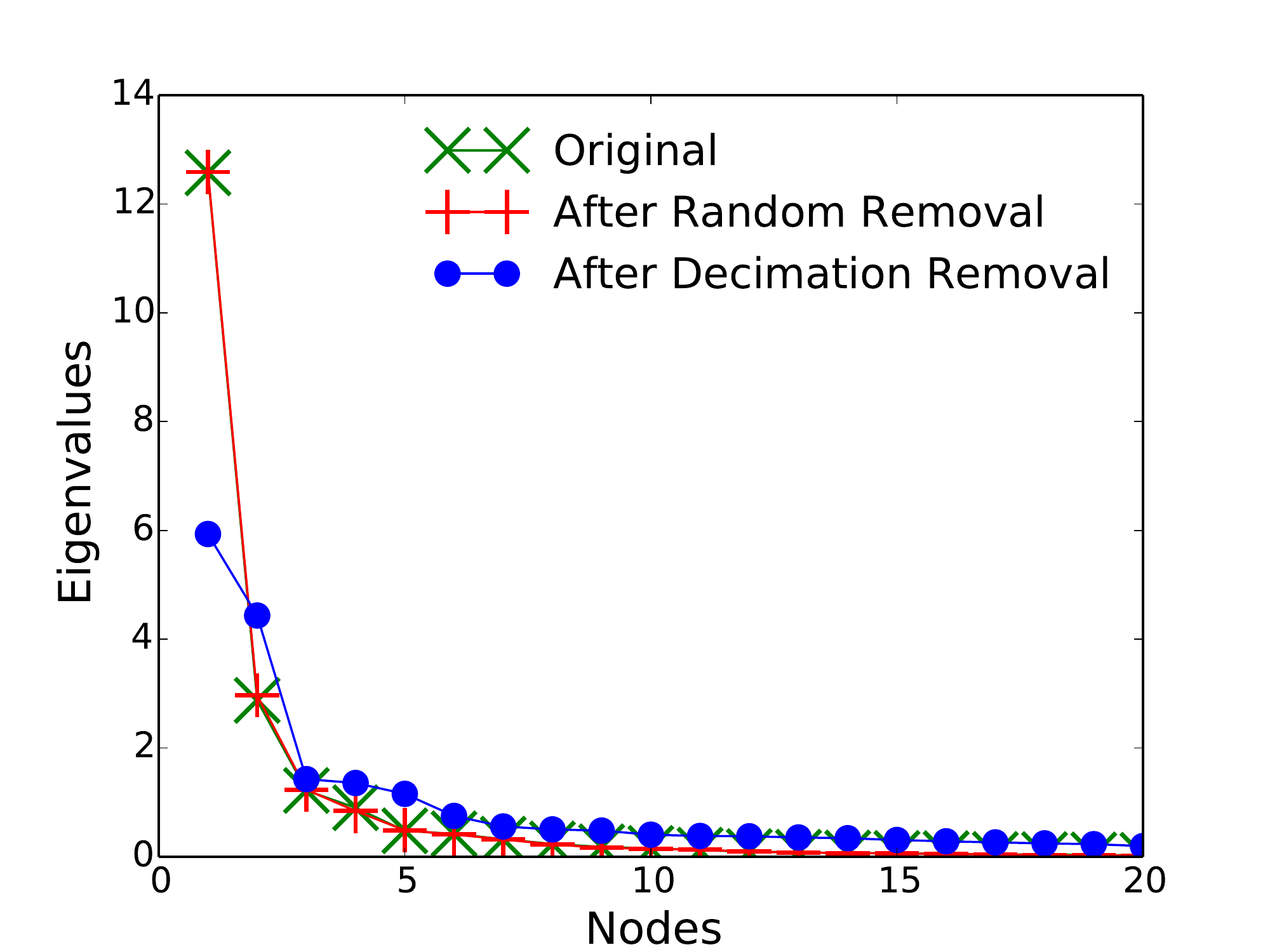}
   \caption{\textit{Top}: Evolution of the effective rank of the correlation matrix $C$,
   $R_\text{entro}$ (\textit{left}) and $R_\text{IPR}$ (\textit{right}), with the fraction of
   configurations removed randomly and using the decimation algorithm.
	\textit{Bottom left}: Reconstruction error obtained using as input the decimated dataset.
	\textit{Bottom right}: Eigenvalues sorted in decreasing order, for the original correlation
	matrix and for the correlation matrix after random removal and decimation-based-removal
	of $90\%$ of configurations.
	In all of the figures, $\beta=2.2, T=32$, at each step of the decimation the configuration that 
	gives the largest increase in the effective rank among $100$ randomly sampled
	configurations is removed from the dataset.
	\label{fig:dec}}
\end{figure}

\begin{figure}
   \centering
   \includegraphics[width=0.7\columnwidth]{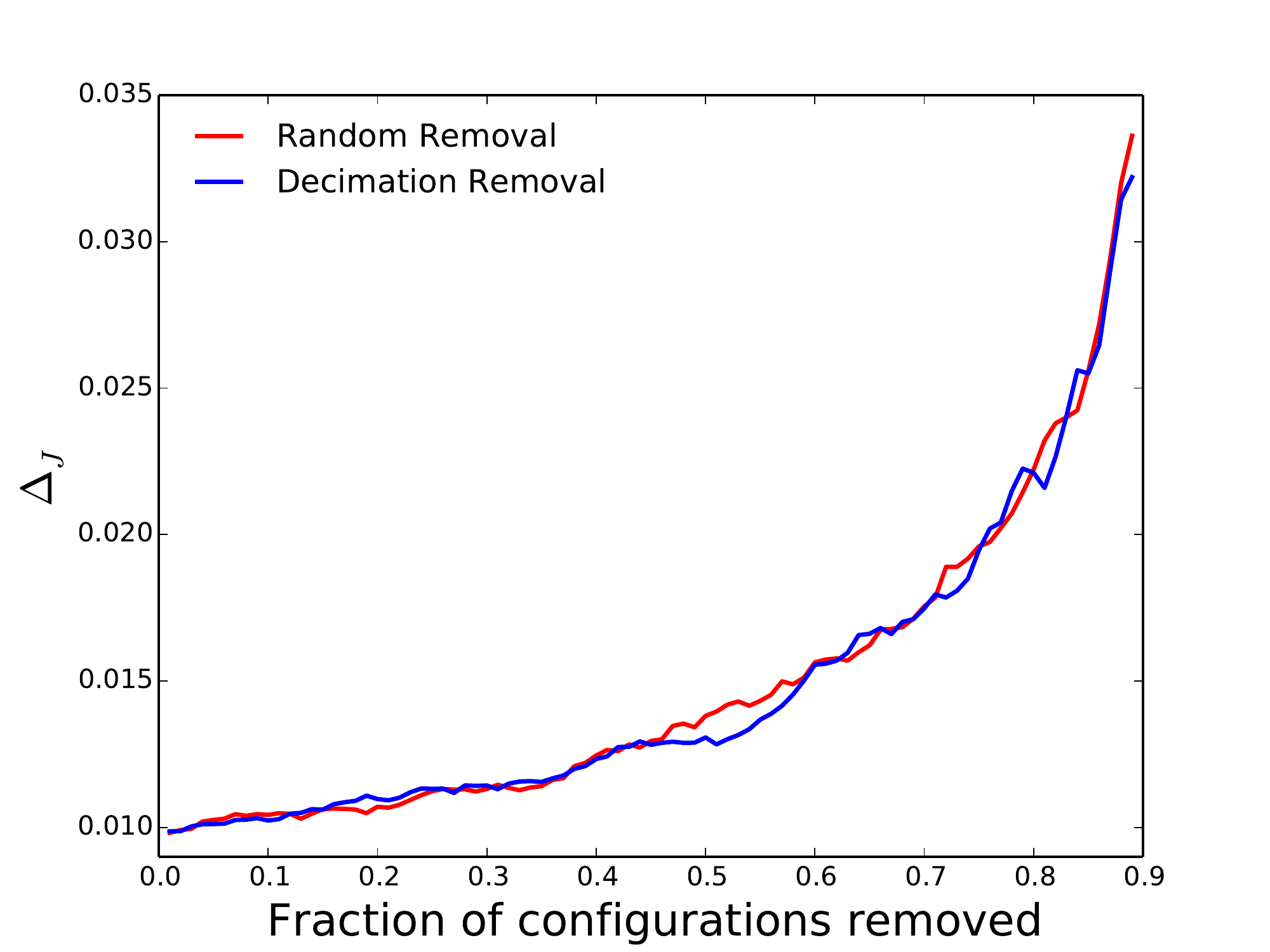}
   \caption{Reconstruction error obtained using as input the decimated dataset, as a function of fraction of edges removed, $1-\alpha$.
   The parameters are the same as Fig.~\ref{fig:dec}, but with $T=0$, and the error grows as $1/\sqrt{\alpha}$.
\label{fig:dec:t0}}
\end{figure}

In the upper panels of Fig.~\ref{fig:dec} we plot the evolution of the effective rank as a function of fraction of configurations removed, for the decimation method just described (blue line) and for the process of removing randomly chosen configurations (red line). We clearly see that the choice based on Eq.~\eqref{eq:deltaR} leads to an increase of the effective rank and consequently of the data quality, while the random decimation keeps the effective rank roughly unchanged (actually there is a small, but systematic decrease; notice that $\Delta R$ can be negative).
As a consequence, if we infer the couplings starting from the decimated dataset, whose size is $\alpha$ times the original one, we obtain a reconstruction error that grows as $1/\sqrt{\alpha}$ for the random decimation (see red line in the lower left panel of Fig.~\ref{fig:dec}). On the contrary, the dataset decimated according to our new rule returns a much smaller reconstruction error. For example, we are able to reduce by a factor 10 the size of the dataset by increasing by less than a factor 2 the reconstruction error (see blue line in the lower left panel of Fig.~\ref{fig:dec}).
The lower right panel of Fig.~\ref{fig:dec} shows the eigenvalues of the $C$ matrix when the dataset is reduced by a factor 10 ($\alpha=0.1$). The random decimation process keeps the eigenvalues practically unchanged, while our decimation algorithm strongly reduces the largest eigenvalues, thus decreasing the redundancy of the dataset.

As a comparison, in Fig.~\ref{fig:dec:t0} we show the reconstruction
error as a function of the fraction of edge removed for the same network
used in Fig.~\ref{fig:dec}, but with $T=0$. In this case, configurations
are randomly chosen and we see that decimation is not useful
anymore, as there is no redundancy in the data, as opposed to the 
equilibrium data.

\section{From one-step to multi-step experiments}
\label{sec:multi}
In previous sections we have discussed the one-step experiments. In this
section we consider the case that measurements are taken from 
a longer trajectory of $L+1$ steps. Although we have mentioned earlier that
a trajectory of length $L+1$ can always be seen as the union of $L$ one-step experiments,
we would like to show that in the multi-step settings ($L>1$) our results
still hold.

\begin{figure}
   \centering
   \includegraphics[width=0.48\columnwidth]{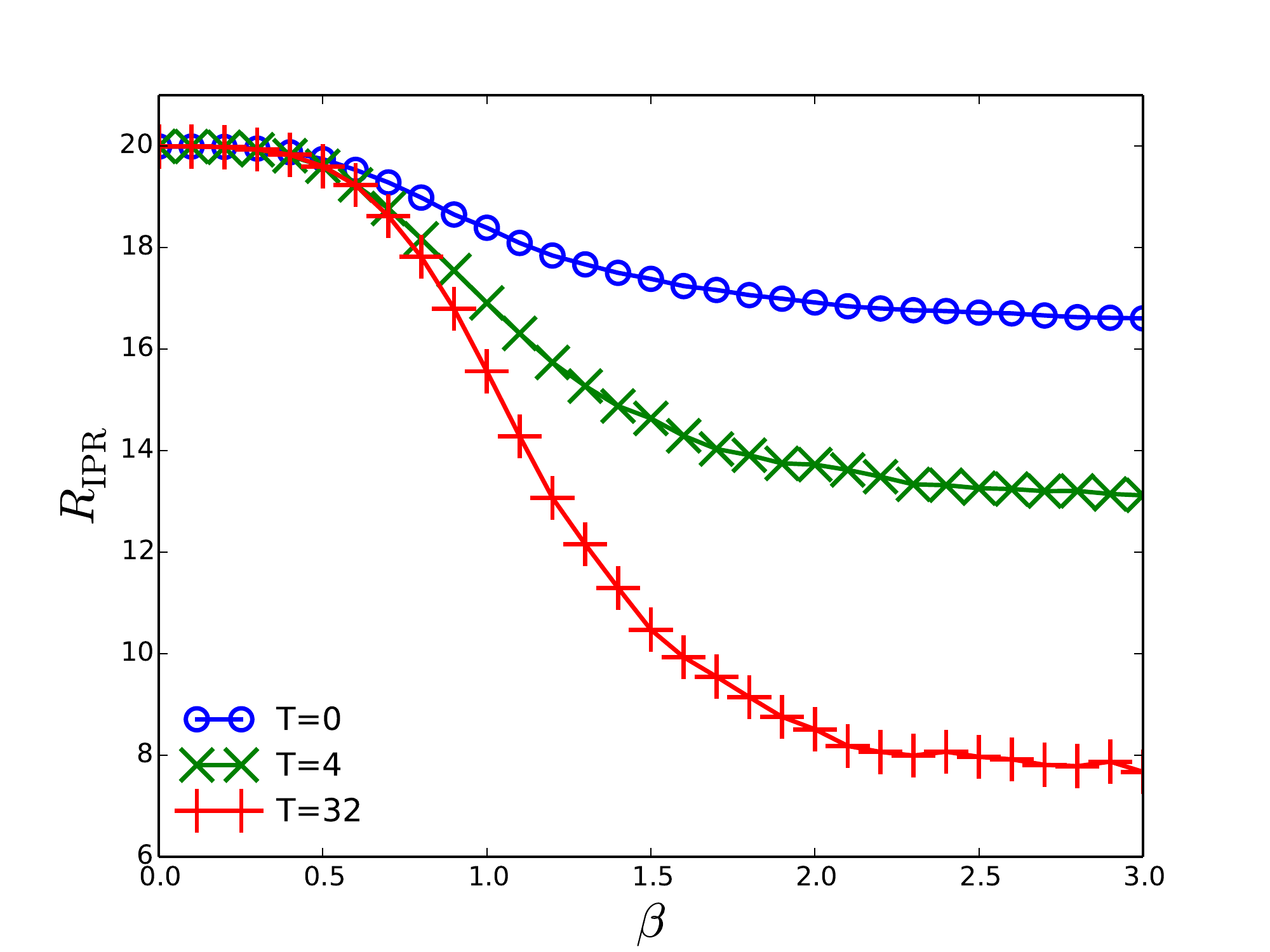}
   \includegraphics[width=0.48\columnwidth]{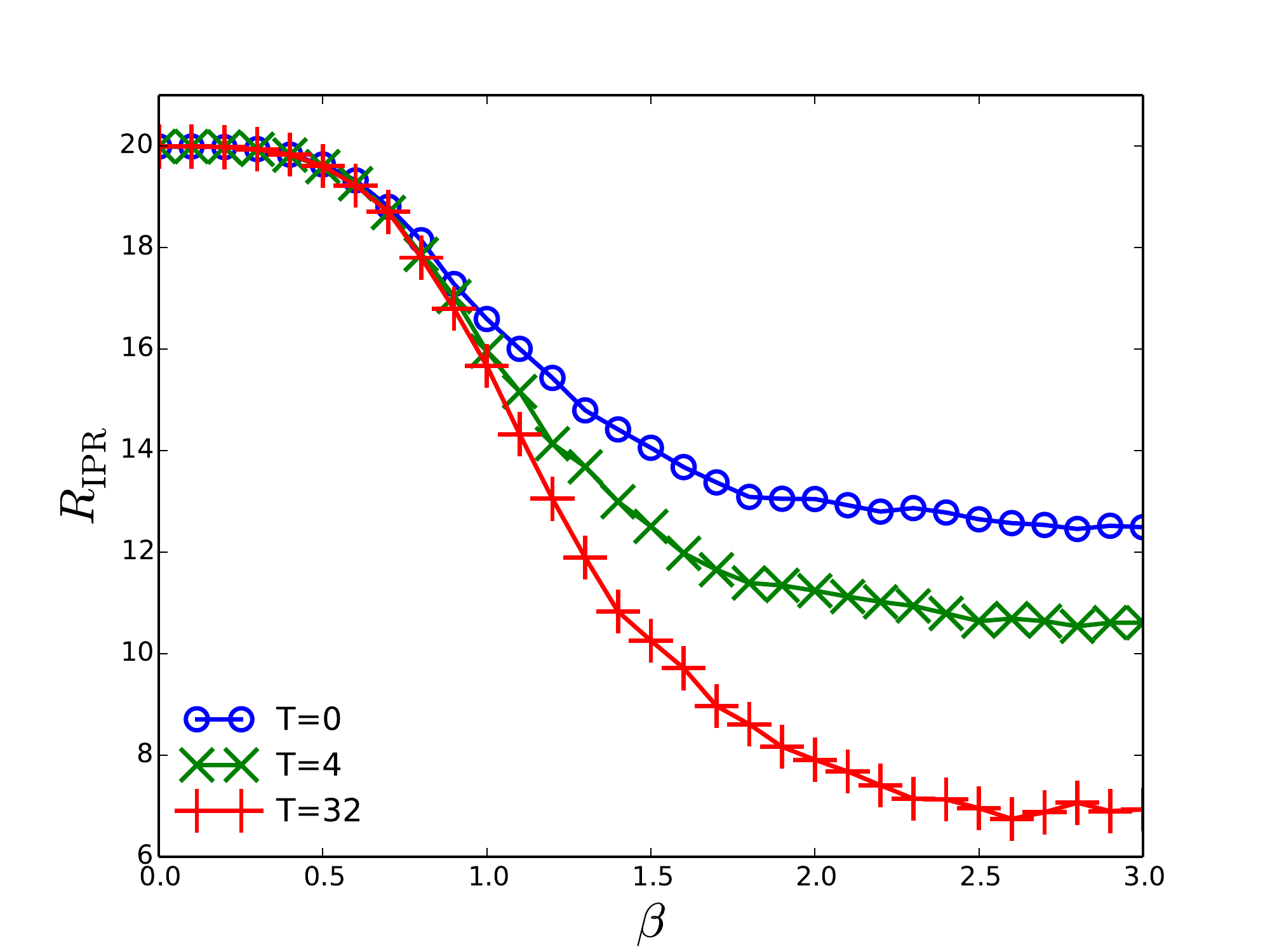}
   \includegraphics[width=0.48\columnwidth]{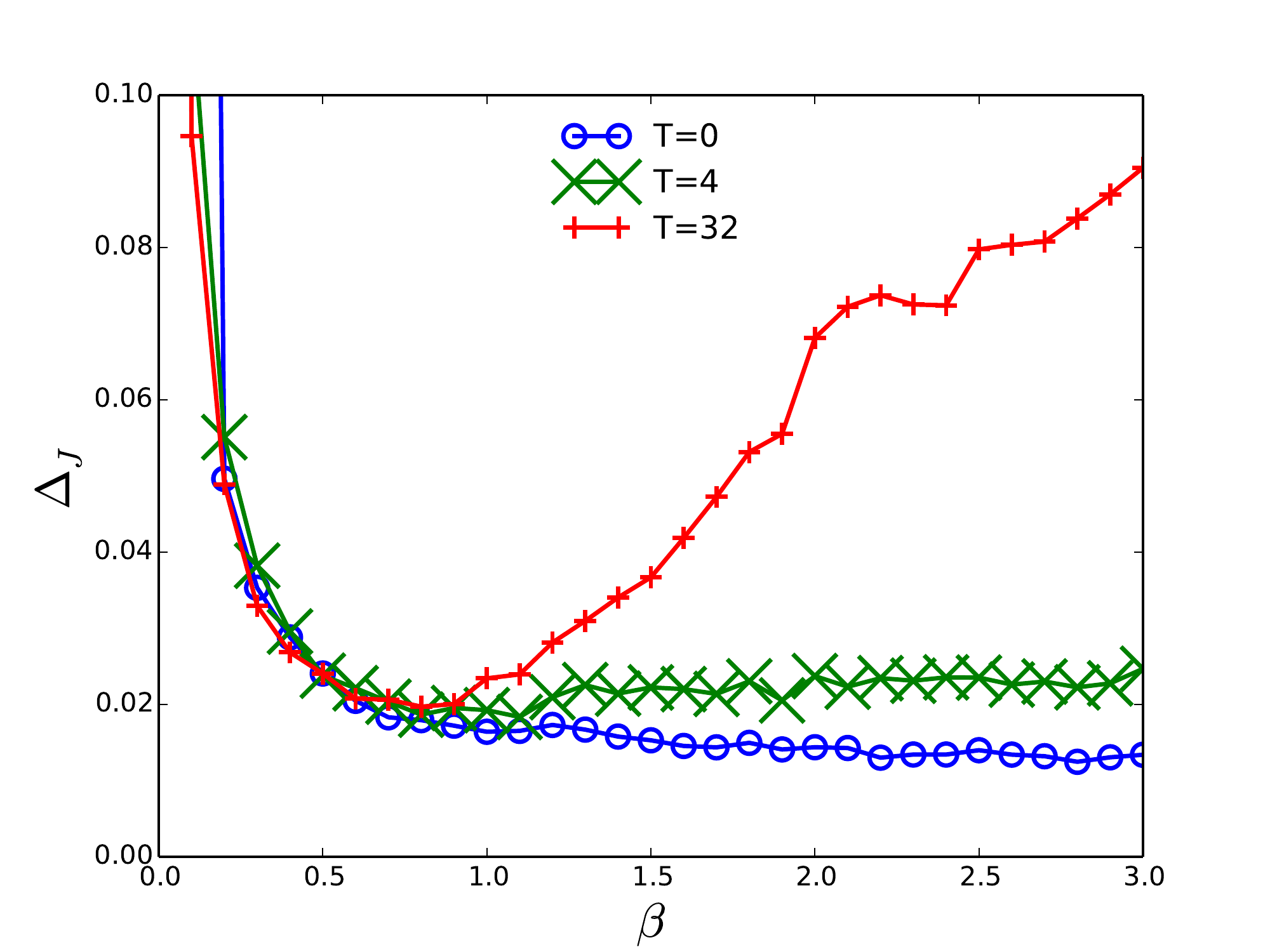}
   \includegraphics[width=0.48\columnwidth]{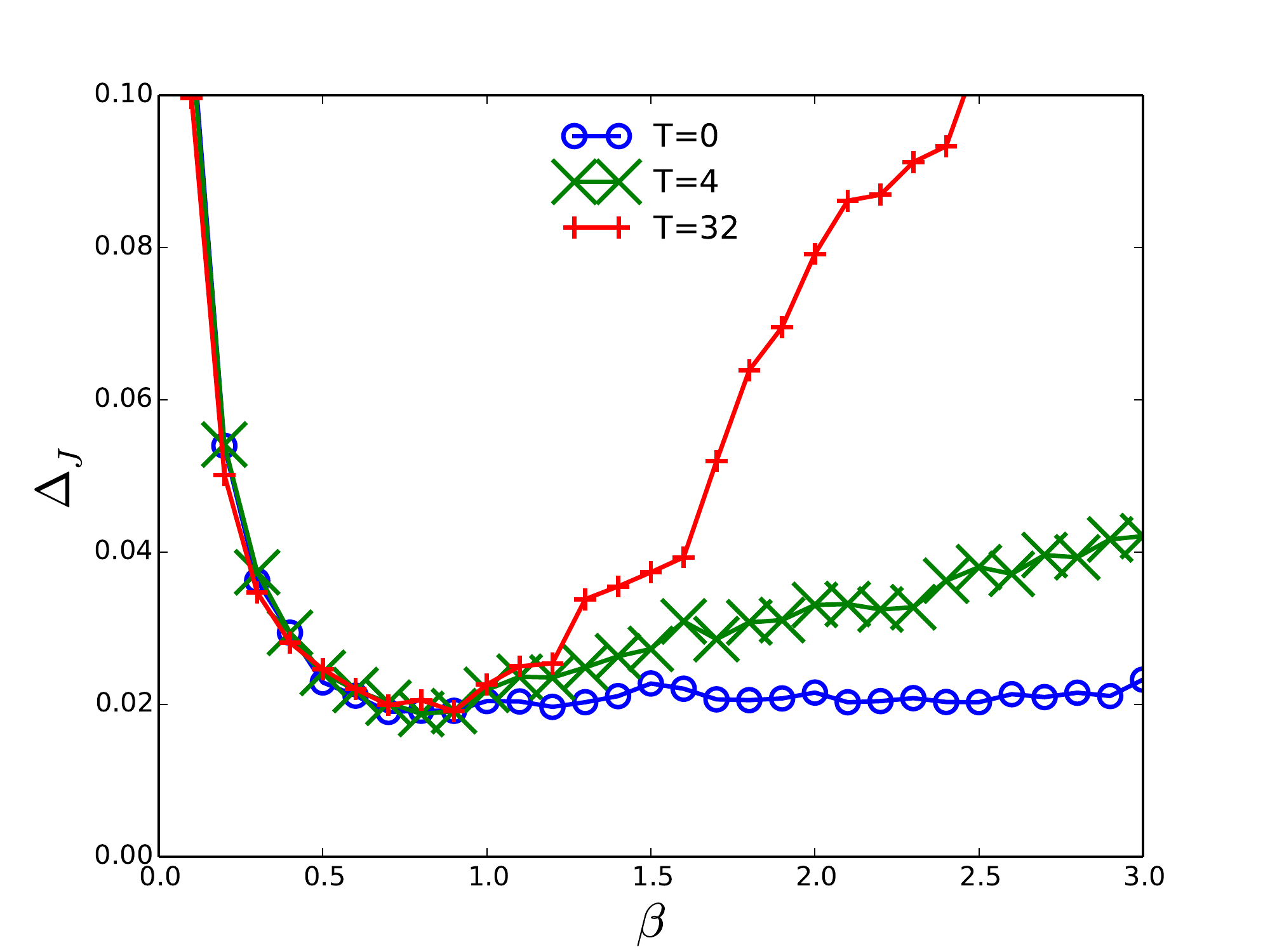}
   \caption{\textit{Top panels}: The effective rank of the correlation matrix $C$,
   $R_\text{IPR}$(\textit{top panels}) and inference error (\textit{bottom panels}) as
   a function of $\beta$ for $L=8$ (\textit{left panels}) and $L=32$ (\textit{right panels}),
   as a function of $\beta$. In experiments, the network 
   is the same as Fig.~\ref{fig:entropy}, which has $n=20$ spins and different $T$,
   using $m=10^4$ configurations.
	\label{fig:L}}
\end{figure}

Given $m$ trajectories with $t\in[T,T+L]$ we maximize the log-likelihood
\begin{align}
    \mathcal{L} & = \sum_{t=T+1}^{T+L} \bigg[ \sum_{i,j} \beta J_{ij} \langle s_i^{(t)} s_j^{(t-1)} \rangle_\data - \sum_i \langle \log( 2 \cosh(\beta \sum_j J_{ij} s_j^{(t-1)}) ) \rangle_\data\bigg],
\end{align}
with respect to the couplings $\{J_{ij}\}$ to be inferred.
In Fig.~\ref{fig:L} we show the effective rank and inference
error as a function of $\beta$ on the same network used in 
Fig.~\ref{fig:entropy}, but for two different $L$ values.
(notice that the previous results can be seen as the special case with $L=1$).
We see that the results in Fig.~\ref{fig:L} are analogous to those in Fig.~\ref{fig:entropy} and
Fig.~\ref{fig:pca}: that is for $\beta > 1$, a larger $T$ value corresponds to worse data quality
and larger inference error. Moreover notice that with $L>1$, it is not possible
to gather data in a purely out-of-equilibrium regime. Indeed 
we can see from the figure that the larger $L$, 
the more equilibrated the data is, resulting in a worse data quality.

A completely different situation would arise if only the first ($t=T$) and the last ($t=T+L$) configurations in a length $L$ trajectory were available. In this case all variables with time index $t\in[T+1,T+L-1]$ are hidden, i.e.\ not directly observed, and the inference problem is much harder (see e.g.\ Ref.~\cite{marsili2013sampling} and references therein). This harder problem is out of the scope of the present work, because the larger uncertainty introduced by the presence of hidden variables would make the connection between data quality and inference error weaker.

\section{Data quality in the static inverse Ising problem}
Although the problem discussed above uses transient dynamics as measurements to infer the
underlying couplings, our analysis also applies to the static problem, the so-called Boltzmann machine learning problem (without hidden variables).
In the static problem, the task is again to infer couplings from a
set of configurations that were generated from the model.
However rather than acquired from transient dynamics of the 
model, in the static case the $m$ configurations are sampled from the Boltzmann distribution:
\begin{equation}
	P(\underline s)=\frac{1}{Z}e^{\beta \sum_{\brc{ij}} J_{ij}s_is_j},
\end{equation}
with partition function 
$$Z=\sum_{\{\underline s\}}e^{\beta \sum_{\brc{ij}} J_{ij}s_is_j}.$$
Thus the couplings can be reconstructed by maximizing the likelihood of
the model
\begin{equation}
	\mathcal{L}(\{\underline s^{(1)},\underline s^{(2)},...,\underline s^{(m)}\})=\sum_{t=1}^m
	\sum_{\brc{ij}}\beta J_{ij}s^{(t)}_is_j^{(t)}-m\log Z.
\end{equation}
However, in the last equation $\log Z$ is difficult to compute, so
authors in ~\cite{aurell2012inverse} introduced the pseudo-likelihood
which approximates the joint probability of a configuration using the 
product of conditional probabilities:
\begin{align}
	\mathcal{L}_p(\{\underline s^{(1)},\underline s^{(2)},...,\underline s^{(m)}\})&=
	\sum_{t=1}^m\sum_{i=1}^n\log P(s_i^{(t)}|\underline s^{(t)}),
\end{align}
here
\begin{equation}
	P(s_i^{(t)}|\underline s^{(t)})=\frac{e^{\beta s_i^{(t)}\sum_{j\neq i}J_{ij}s_j^{(t)}}}{2\cosh(\sum_{j\neq i}J_{ij}s_j^{(t)})}.
\end{equation}
If we put configurations $\{\underline s^{(1)},\underline s^{(2)},...,\underline s^{(m)}\}$ into rows of matrix $A$, as we did for the dynamical case, the log-pseudo-likelihood can be written as
\begin{align}
	\label{eq:pl}
\mathcal{L}_p  = \sum_{a=1}^{m}\brb{\sum_{i,j} \beta J_{ij} A_{ai}A_{aj} - \sum_i \log( 2 \cosh(\beta \sum_j J_{ij} A_{aj}) )}.
\end{align}
We see that the last equation has the same form as Eq.~\eqref{eq:likelihood}. 
The only difference is that matrix $B$ is replaced by $A$, that is, time
indices are different.

\begin{figure}
   \centering
   \includegraphics[width=0.45\linewidth]{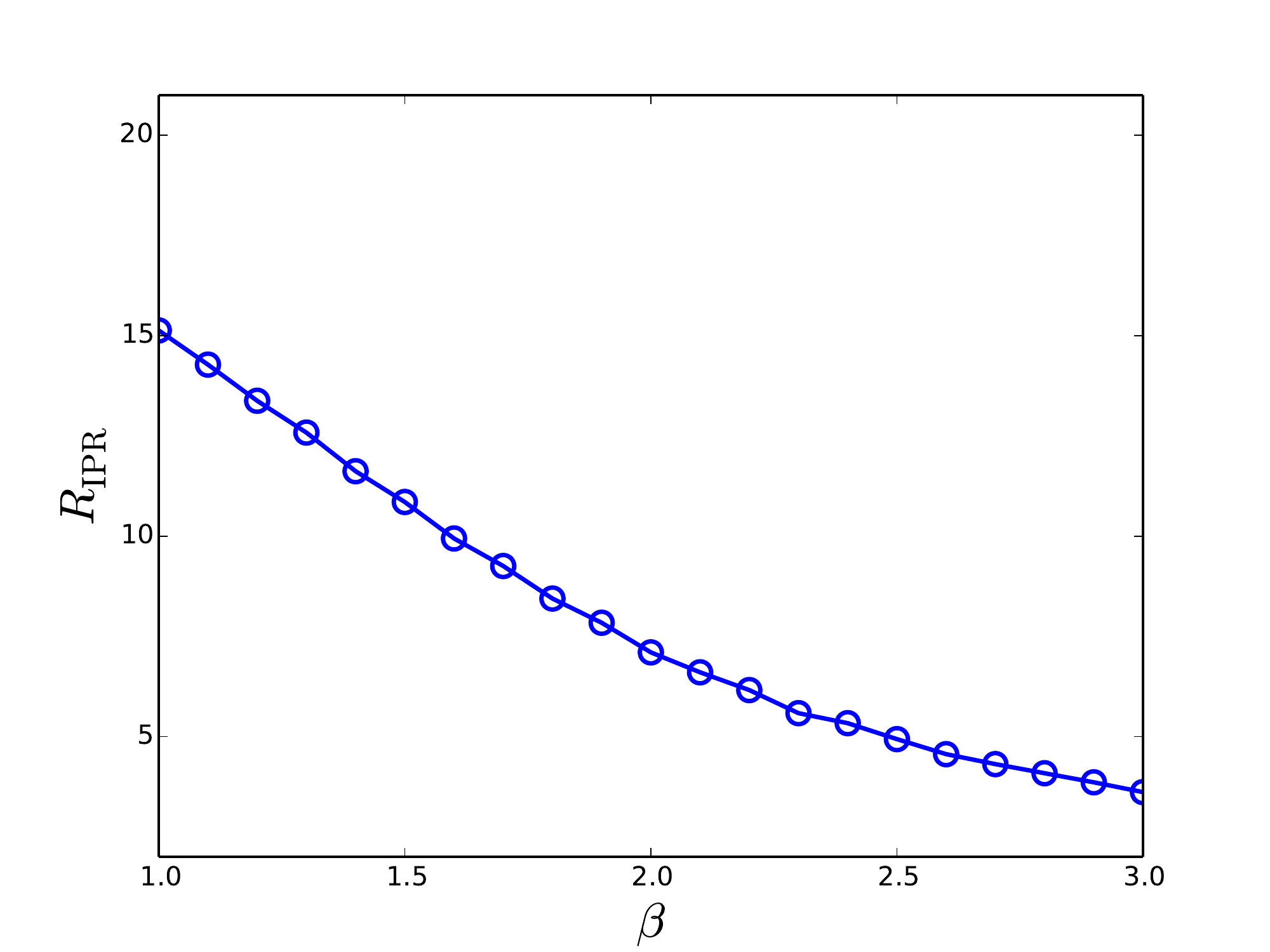}
   \includegraphics[width=0.45\linewidth]{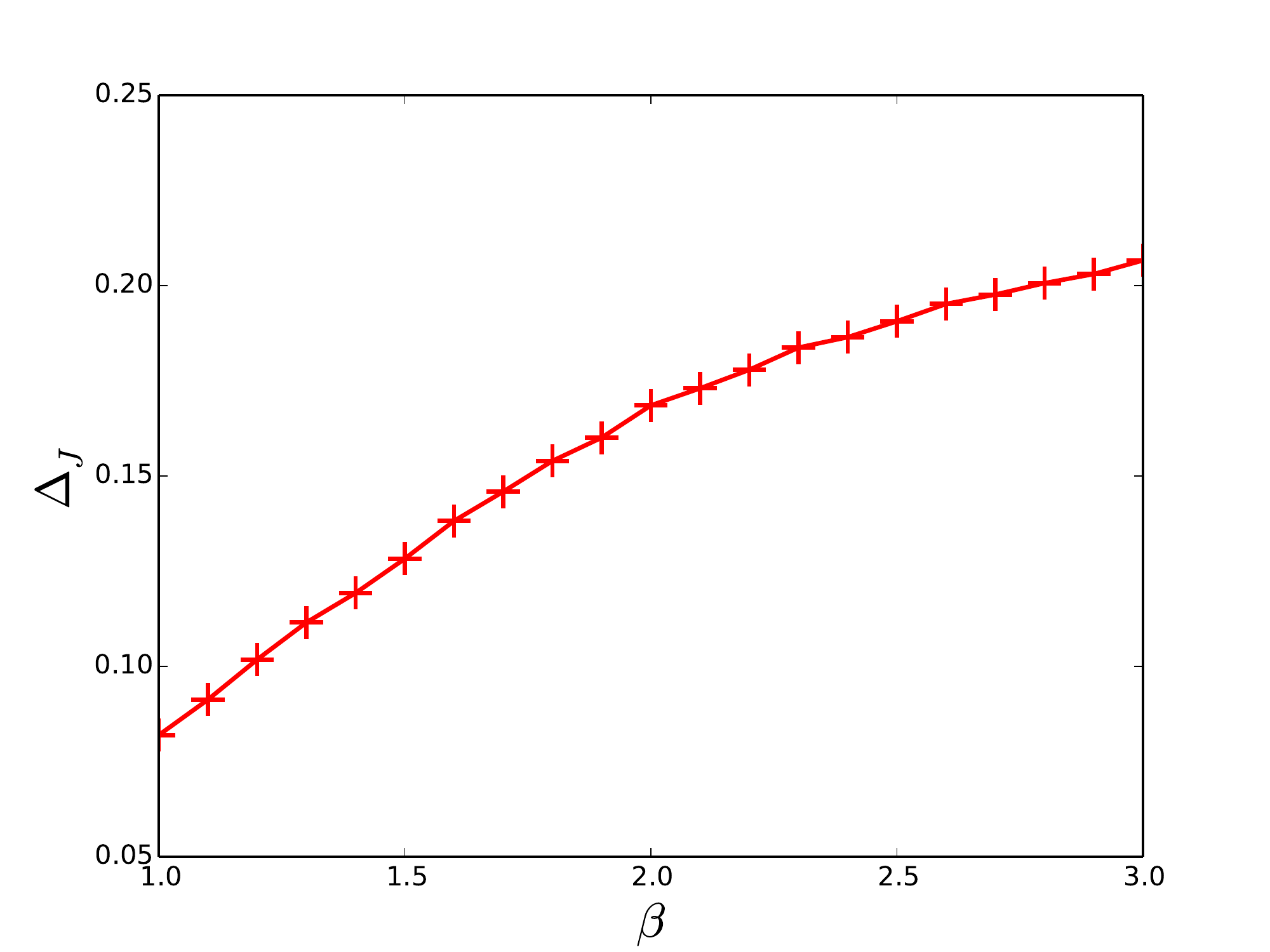}
   \caption{Effective rank $R_{\textrm{IPR}}$ (\textit{left}) and 
   inference error by maximizing the pseudo likelihood Eq.~\eqref{eq:pl}
   for a static inverse Ising model. The model is fully connected SK 
   model with $n=20$ spins, using $m=10^4$ configurations sampled from
   the Boltzmann distribution.
   \label{fig:pl}}
\end{figure}

So our data-quality analysis targeting matrix $A$, made above for the dynamical model,
can be applied directly to the 
the pseudo-likelihood-based inference of the static model. 
In Fig.\ref{fig:pl} we plot the effective rank $R_{\textrm{ipr}}$ and
the inference error by maximizing the pseudo likelihood 
Eq.~\eqref{eq:pl} for a static inverse Ising model. We see that
the result is analogous to the dynamical case: a larger effective
rank results to a smaller inference error, revealing a better quality of data.

We note that both the dynamical model and the pseudo-likelihood based 
static model belong to the class of generalized linear inference 
model with effectively the logistic function as a kernel 
(as in the logistic regression). But keep in mind that
they have specific data (configurations) that are generated by a physical model, 
the Ising model, and the analysis of the quality of data from the 
physical model, is indeed our focus in this paper.

\section{Conclusion and discussion}
\label{sec:con}

We have studied the data quality problem in the kinetic inverse Ising problem. 
First, we have experimentally shown that data gathered in an out-of-equilibrium regime has 
better quality and thus leads to a smaller reconstruction error than data gathered in equilibrium.
Then we focused on how to quantify the data quality using the effective 
rank of the correlation matrix, and how to 
improve the data quality by a decimation procedure based on a perturbative analysis of 
the correlation matrix.


Though we only studied the SK model in this paper, we have tested other disordered
models such as the Hopfield model and the sequence processing neural networks, where the
results are qualitatively similar.

In all of the experiments, the inference is done by maximizing the likelihood, which
is usually prone to overfitting, especially when the number of configurations is not
large enough. It would be interesting to extend our results and the decimation method
to Bayesian inference of the model parameters. We leave this for future work.

We believe the results of the present work can be very useful in applications. On the one hand, we have shown that data collected in a strongly out-of-equilibrium regime are much more informative about the interaction network of a set of dynamically interacting variables, and this may suggest new ways to collect the data to infer such an interaction network: for example, perturbing the system out of its equilibrium/stationary regime may allow the system to show up more clearly correlations and fluctuations, that are useful for the reconstruction problem.
On the other hand, given that the amount of data available is growing very fast in recent years, the decimation method we have proposed for strongly reducing the size of the input dataset, without losing too much information for the reconstruction problem, may be extremely practical for dealing with huge datasets. Notice that the method can be used also on-the-run, i.e.\ while data are being generated: in this case one can accept only configurations that bring a substantial improvement in the effective rank, and leaving aside redundant data.

\section*{References}

\bibliographystyle{unsrt}


\begin{thebibliography}{10}

\bibitem{ackley1985learning}
David~H Ackley, Geoffrey~E Hinton, and Terrence~J Sejnowski.
\newblock A learning algorithm for boltzmann machines.
\newblock {\em Cognitive science}, 9(1):147--169, 1985.

\bibitem{Schneidman_etal_2006_nature}
Elad Schneidman, Michael~J. Berry, Ronen Segev, and William Bialek.
\newblock Weak pairwise correlations imply strongly correlated network states
  in a neural population.
\newblock {\em Nature}, 440:1007--1012, 2006.

\bibitem{cocco2011adaptive}
Simona Cocco and R{\'e}mi Monasson.
\newblock Adaptive cluster expansion for inferring boltzmann machines with
  noisy data.
\newblock {\em Physical review letters}, 106(9):90601, 2011.

\bibitem{ricci2011mean}
Federico Ricci-Tersenghi.
\newblock The bethe approximation for solving the inverse ising problem: a
  comparison with other inference methods.
\newblock {\em J. Stat. Mech.}, 2012(08):P08015, 2012.

\bibitem{inf_nguyen-11}
H.~Chau Nguyen and J.~Berg.
\newblock Bethe-peierls approximation and the inverse ising model.
\newblock {\em arXiv}, 1112.3501, 2011.

\bibitem{inf_nguyen-12}
H.~Chau Nguyen and J.~Berg.
\newblock Mean-field theory for the inverse ising problem at low temperatures.
\newblock {\em arXiv}, 1204.5375, 2012.

\bibitem{aurell2012inverse}
Erik Aurell and Magnus Ekeberg.
\newblock Inverse ising inference using all the data.
\newblock {\em Physical review letters}, 108(9):90201, 2012.

\bibitem{zhang12}
Pan Zhang.
\newblock Inference of kinetic ising model on sparse graphs.
\newblock {\em Journal of Statistical Physics}, 148(3):502--512, 2012.

\bibitem{decelle13}
Aur\'elien Decelle and Federico Ricci-Tersenghi.
\newblock Pseudolikelihood decimation algorithm improving the inference of the
  interaction network in a general class of ising models.
\newblock {\em Phys. Rev. Lett.}, 112:070603, Feb 2014.

\bibitem{wainwright2010AnnalHigh}
B.~Ravikumar, M.J. Wainwright, and J.D. Lafferty.
\newblock High-dimensional ising model selection using l1-regularized logistic
  regression.
\newblock {\em The Annals of Statistics}, 38, 2010.

\bibitem{roudi2009}
Yasser Roudi, Joanna Tyrcha, and John Hertz.
\newblock Ising model for neural data: Model quality and approximate methods
  for extracting functional connectivity.
\newblock {\em Phys. Rev. E}, 79:051915, May 2009.

\bibitem{fortunato2010community}
Santo Fortunato.
\newblock Community detection in graphs.
\newblock {\em Physics Reports}, 486(3):75--174, 2010.

\bibitem{ruz2010learning}
Gonzalo~A Ruz and Eric Goles.
\newblock Learning gene regulatory networks with predefined attractors for
  sequential updating schemes using simulated annealing.
\newblock In {\em Machine Learning and Applications (ICMLA), 2010 Ninth
  International Conference on}, pages 889--894. IEEE, 2010.

\bibitem{weigtPNAS}
Martin Weigt, Robert~A White, Hendrik Szurmant, James~A Hoch, and Terence Hwa.
\newblock Identification of direct residue contacts in protein--protein
  interaction by message passing.
\newblock {\em Proceedings of the National Academy of Sciences}, 106(1):67--72,
  2009.

\bibitem{ekeberg2013improved}
Magnus Ekeberg, Cecilia L{\"o}vkvist, Yueheng Lan, Martin Weigt, and Erik
  Aurell.
\newblock Improved contact prediction in proteins: Using pseudolikelihoods to
  infer potts models.
\newblock {\em Physical Review E}, 87(1):012707, 2013.

\bibitem{bialek2012statistical}
William Bialek, Andrea Cavagna, Irene Giardina, Thierry Mora, Edmondo
  Silvestri, Massimiliano Viale, and Aleksandra~M Walczak.
\newblock Statistical mechanics for natural flocks of birds.
\newblock {\em Proceedings of the National Academy of Sciences},
  109(13):4786--4791, 2012.

\bibitem{decelle15}
Aur\'elien Decelle and Pan Zhang.
\newblock Inference of the sparse kinetic ising model using the decimation
  method.
\newblock {\em Phys. Rev. E}, 91:052136, May 2015.

\bibitem{ferrari2015approximated}
Ulisse Ferrari.
\newblock Approximated newton algorithm for the ising model inference speeds up
  convergence, performs optimally and avoids over-fitting.
\newblock {\em arXiv preprint arXiv:1507.04254}, 2015.

\bibitem{raymond2013mean}
Jack Raymond and Federico Ricci-Tersenghi.
\newblock Mean-field method with correlations determined by linear response.
\newblock {\em Physical Review E}, 87(5):052111, 2013.

\bibitem{roudi2011mean}
Yasser Roudi and John Hertz.
\newblock Mean field theory for nonequilibrium network reconstruction.
\newblock {\em Physical review letters}, 106(4):048702, 2011.

\bibitem{mezard2011exact}
M~M{\'e}zard and J~Sakellariou.
\newblock Exact mean-field inference in asymmetric kinetic ising systems.
\newblock {\em Journal of Statistical Mechanics: Theory and Experiment},
  2011(07):L07001, 2011.

\bibitem{zeng2011network}
Hong-Li Zeng, Erik Aurell, Mikko Alava, and Hamed Mahmoudi.
\newblock Network inference using asynchronously updated kinetic ising model.
\newblock {\em Physical Review E}, 83(4):041135, 2011.

\bibitem{zeng2013maximum}
Hong-Li Zeng, Mikko Alava, Erik Aurell, John Hertz, and Yasser Roudi.
\newblock Maximum likelihood reconstruction for ising models with asynchronous
  updates.
\newblock {\em Physical review letters}, 110(21):210601, 2013.

\bibitem{bachschmid2015learning}
Ludovica Bachschmid-Romano and Manfred Opper.
\newblock Learning of couplings for random asymmetric kinetic ising models
  revisited: random correlation matrices and learning curves.
\newblock {\em Journal of Statistical Mechanics: Theory and Experiment},
  2015(9):P09016, 2015.

\bibitem{sherrington1975solvable}
David Sherrington and Scott Kirkpatrick.
\newblock Solvable model of a spin-glass.
\newblock {\em Physical review letters}, 35(26):1792, 1975.

\bibitem{roy2007effective}
Olivier Roy and Martin Vetterli.
\newblock The effective rank: A measure of effective dimensionality.
\newblock In {\em European signal processing conference (EUSIPCO)}, number
  LCAV-CONF-2007-017, pages 606--610, 2007.

\bibitem{Mezard2002}
M.~M\'ezard, G.~Parisi, and R.~Zecchina.
\newblock Analytic and algorithmic solution of random satisfiability problems.
\newblock {\em Science}, 297:812, 2002.

\bibitem{marsili2013sampling}
Matteo Marsili, Iacopo Mastromatteo, and Yasser Roudi.
\newblock On sampling and modeling complex systems.
\newblock {\em Journal of Statistical Mechanics: Theory and Experiment},
  2013(09):P09003, 2013.

\end{thebibliography}


\end{document}